\documentclass[reqno]{amsart}
\usepackage{graphicx}
\usepackage{epstopdf, epsfig}
\usepackage{subeqnarray}

\usepackage{tikz}
\usetikzlibrary{decorations.pathmorphing}
\usepackage{amsmath}

\usepackage{graphicx}
\usepackage{subcaption}
\usepackage{amssymb,amsfonts}
\usepackage{floatrow}
\usepackage{enumerate}
\usepackage{bm}
\usepackage{natbib}
\usepackage[margin=1in]{geometry}
\usepackage{hyperref}

\makeatletter
\newsavebox{\@brx}
\newcommand{\llangle}[1][]{\savebox{\@brx}{\(\m@th{#1\langle}\)}%
  \mathopen{\copy\@brx\kern-0.5\wd\@brx\usebox{\@brx}}}
\newcommand{\rrangle}[1][]{\savebox{\@brx}{\(\m@th{#1\rangle}\)}%
  \mathclose{\copy\@brx\kern-0.5\wd\@brx\usebox{\@brx}}}
\makeatother

\newcommand{\Weber}{\operatorname{\mathit{W\kern-.30em e}}}
\newcommand{\Rey}{\operatorname{\mathit{R\kern-.20em e}}}
\newcommand{\Capil}{\operatorname{\mathit{C}}}

\title{Three-dimensional wave evolution on electrified falling films}

\begin{document}

\author{R. J. Tomlin,
 D. T. Papageorgiou
\and
  G. A. Pavliotis
}

\begin{abstract}
We consider the full three-dimensional dynamics of a thin falling liquid film on a flat plate inclined at some non-zero angle to the horizontal. In addition to gravitational effects, the flow is driven by an electric field which is normal to the substrate far from the flow. This extends on the work of Tseluiko and Papageorgiou \cite{tseluiko2006wave} by including transverse dynamics. We study both the cases of overlying and hanging films, where the liquid rests above and below the substrate respectively. Starting with the Navier--Stokes equations coupled with electrostatics, a fully nonlinear two-dimensional Benney equation for the interfacial dynamics is derived valid for waves that are long compared to the film thickness. The weakly nonlinear evolution is found to be governed by a Kuramoto--Sivashinsky equation with a non-local term due to the electric field effect. The electric field term is linearly destabilising and produces growth rates proportional to $|\bm{\xi}|^3$, where $\bm{\xi}$ is the wavenumber vector of the perturbations. It is found that transverse gravitational instabilities are always present for hanging films and this leads to unboundedness of nonlinear solutions even in the absence of electric fields - this is due to the isotropy of the nonlinearity in the flow direction. For overlying films and a restriction on the strength of the electric field, the equation is well-posed in the sense that it possesses bounded solutions. This two-dimensional equation is studied numerically for the case of periodic boundary conditions in order to assess the effects of inertia, electric field strength the dimensions of the periodic domain. Rich dynamical behaviours are observed and classified in various parameter windows. For subcritical Reynolds number flows, a sufficiently strong electric field can promote non-trivial dynamics for some choices of domain dimensions, leading to fully two-dimensional evolutions for the interface. These dynamics are also found to produce two-dimensional spatiotemporal chaos on sufficiently large domains. For supercritical flows, such two-dimensional chaotic dynamics emerge in the absence of a field, and its presence enhances the amplitude of the fluctuations and broadens their spectrum.
\end{abstract}

\maketitle


\section{Introduction\label{IntroductionSec}}

Thin liquid films arise in many physical applications, in particular cooling and coating processes. In the case of cooling, numerous studies \cite{shmerler1986effects,LYU19911451,Miyara1999,serifi2004transient,Aktershev2010,Aktershev2013,Mascarenhas20131106} showed evidence that interfacial waves increase heat transfer by orders of magnitude. This phenomenon was shown to be caused by convection effects and film thinning. 
For coating processes however, a stable thin film of relatively constant thickness is required to evenly coat the surface of a substrate. It was shown by Benjamin \cite{FLM:367246} and Yih \cite{Yih1} that there is a critical Reynolds number depending on the angle of inclination, above which a thin film becomes unstable to long waves. For Reynolds numbers close to this critical value, it is viable to use long-wave asymptiotics to produce a nonlinear Benney equation to describe the interface evolution \cite{Benney}. The addition of an electric field to the thin film flow problem gives rise to additional stresses acting at the fluid interface which in turn
affect the flow stability; electric fields can promote non-trivial dynamics for flows that would be stable in their absence. Melcher and Taylor \cite{MelcherTaylor} reviewed the early work on the modelling of perfectly conducting liquids and perfect dielectrics, and developed the Taylor--Melcher leaky dielectric model for poorly conducting fluids which was then studied extensively \cite{feng1996computational, doi:10.1146/annurev.fluid.29.1.27}, even in the thin film context \cite{pease2002linear,:/content/aip/journal/pof2/17/3/10.1063/1.1852459}. The possibility of controlling film flows using vertical electric fields was considered by a number of authors \cite{kim1992effect,kim1994cylindrical,bankoff1994design,bankoff2002use,griffing2006electrohydrodynamics} in their study of the electrostatic liquid--film radiator.

The two-dimensional simplification of our model - yielding one-dimensional evolution equations for the interface - has been studied firstly by Gonz\'alez and Castellanos in \cite{PhysRevE.53.3573} and then extensively by Tseluiko and Papageorgiou \cite{tseluiko2006wave,tseluiko2006global,PhysRevE.82.016322}, in which a normal electric field acts to destabilise the interface of a gravity driven thin film flow, even for subcritical Reynolds number flows. From a fully nonlinear Benney equation for the interface height, they study the weakly nonlinear evolution of the scaled interfacial position $\eta(x,t)$ that satisfies the canonical equation
\begin{equation}\label{KSE1dE1}
\eta_t + \eta \eta_x \pm \eta_{xx} + \gamma \mathcal{H}(\eta_{xxx}) + \eta_{xxxx} = 0,
\end{equation}
where $\mathcal{H}$ is the Hilbert transform and $\gamma\ge0$ measures the strength of the applied
electric field; the $-$ or $+$ is taken depending on whether the Reynolds number is subcritical or supercritical respectively. 
Gonz\'alez and Castellanos identified a critical electric field strength for subcritical Reynolds number flows above which 
instability of a mode with non-zero wavenumber is found and a local bifurcation analysis was performed. 
Tseluiko and Papageorgiou \cite{tseluiko2006wave} completed an extensive numerical study of the initial value problem for (\ref{KSE1dE1}) with periodic boundary conditions on the interval $[0,L]$, finding attractors for the dynamics in windows of the parameters $\gamma$ and $L$. The same authors provide analytical bounds on attractor dimensions and on the solution energy \cite{tseluiko2006global}. The models were extended
to include dispersive effects (expansion to a higher order Benney equation 
is warranted) for the case of vertical film flow \cite{PhysRevE.82.016322}.
Mukhopadhay and Dandapat \cite{mukhopadhyay2004nonlinear} considered the same problem but proceeded with an integral boundary layer formulation, resulting in coupled evolution equations for the fluid flux and interface height. 
Additionally, Tseluiko and Papageorgiou \cite{tseluiko2007nonlinear} studied the case of a horizontal flat substrate by means of long-wave asymptotics for both overlying and hanging films, for a regime in which the capillary number is an order smaller than that of our study, corresponding to a strengthening of surface tension. They provide evidence, using a mixture of numerics and analysis, for the global existence of positive smooth solutions, and furthermore that the film does not touch down at a finite time but approaches the substrate surface asymptotically in infinite time. They also give numerical evidence for this, including the case of
hanging films in the absence of an electric field. 

The present study extends the work described above to fully two-dimensional interfaces. We obtain novel transverse dynamics and show the breakdown of the weakly nonlinear assumption for certain set-ups. We proceed with an analysis similar to that in \cite{tseluiko2006wave} to obtain a fully nonlinear two-dimensional Benney equation for the interface height that retains both inertia and surface tension effects. 
Finite-time blow-up has been observed numerically for the corresponding one-dimensional Benney equation, and in the present work we do not proceed with a numerical study of the two-dimensional Benney equation. 
Instead, we study the weakly nonlinear evolution by perturbing about the exact constant solution for the interface height, 
to obtain a non-local two-dimensional Kuramoto--Sivashinsky-type equation analogous to (\ref{KSE1dE1}). 
Interestingly, the resulting equation is well-posed for overlying films with electric field strengths below a critical value; this is due to transverse instabilities that cannot be saturated
by the nonlinear term. Even in the absence of an electric field, this class of weakly nonlinear
models is not appropriate for the case of hanging films.
For overlying films we will derive the canonical equation
\begin{equation}\label{KSE2dCanon1}\eta_t + \eta \eta_x + (\beta -1) \eta_{xx}  -  \eta_{yy}  - \gamma  \Delta \mathcal{R}(\eta) + \Delta^2 \eta  = 0,\end{equation}
where $\mathcal{R}$ is a non-local operator corresponding to the electric field effect, $\beta > 0$ is a Reynolds number term measuring inertial effects and $0 \leq \gamma \leq 2$ measures the electric field strength as in \eqref{KSE1dE1} (the latter restriction is imposed to prevent unbounded solutions as mentioned above). When supplemented with periodic boundary conditions on the rectangle $Q=[0,L_1]\times [0,L_2]$, we are left with four parameters governing the dynamical behaviour of solutions. For numerical simulations, we reduce this problem by restricting to square periodic domains, setting $L_1 = L_2 = L$, and study of the dynamical behaviours for various choices of $\beta$, $\gamma$, and $L$. A number of authors \cite{kevrekidis1990back,papageorgiou1,papageorgiou2} explored the attractor windows for the well-known one-dimensional Kuramoto--Sivashinsky equation
\begin{equation}\label{KSE1dIntro}
\eta_t + \eta \eta_x + \eta_{xx} + \eta_{xxxx} = 0,
\end{equation}
on periodic domains of length $L$. Increasing $L$ yields windows of steady attractors, travelling wave attractors, 
time-periodic attractors and period-doubling behaviours among other phenomena. In the majority of the parameter windows, 
the solution profiles are found to have a characteristic cellular form. Chaotic attractors are found for sufficiently large $L$ 
and chaotic behaviour persists for $L$ above a certain threshold. A related equation is the two-dimensional Kuramoto--Sivashinsky equation derived by Nepomnyashchy \cite{Nepo1, Nepo2} for thin film flow down a vertical plane,
\begin{equation}\label{KSE2dNepo}
\eta_t + \eta \eta_x + \eta_{xx} + \Delta^2 \eta = 0.
\end{equation}
The dynamics of solutions to (\ref{KSE2dNepo}) are similar to those observed for (\ref{KSE1dIntro}), 
and solutions in the chaotic regime are found to vary weakly in the transverse direction \cite{RJT2DKSabsorbingball}. 
We find even richer dynamical behaviour for (\ref{KSE2dCanon1}) due to the destabilising electric field, which has no directional preference and provides stronger linear instabilities in the mixed Fourier modes. For subcritical Reynolds number flows, $\beta < 1$, a sufficiently strong electric field is required to promote interfacial waves. We examine the attractor windows for both a small subcritical Reynolds number with $\beta = 0.01$, and a moderate one with $\beta = 0.5$. 
For supercritical Reynolds numbers, $\beta > 1$, we observe the usual Kuramoto--Sivashinsky-type dynamics in the absence of an electric field, however its introduction qualitatively changes the dynamics. For supercritical Reynolds number flows we take $\beta = 2$ and
explore the details of the attractors numerically. 

The structure of the paper is as follows.
Section \ref{PhysicalModelandGoverningEquationsSec} gives the physical model and the full formulation of the problem in dimensional variables. In section \ref{ExactSolutionandDimensionlessEquationsSec}, we give an exact Nusselt solution to the problem and rewrite our equations as perturbations of this exact solution. In section \ref{ALong-WaveEvolutionEquationSec}, we make a long-wave assumption and derive a fully nonlinear Benney equation for the interface height. Section \ref{ANon-localKuramoto-SivashinskyEquationSec} contains the analysis and
computations of the canonical weakly nonlinear evolution equation (\ref{KSE2dCanon1}). Finally, section \ref{ConclusionsandFutureDirectionsSec} contains our conclusions and a discussion.

\section{Physical model and governing equations\label{PhysicalModelandGoverningEquationsSec}}

\begin{figure}
\caption{Schematic of the problem.} \label{Setupdiagram}
\begin{tikzpicture}[scale=1.48]
    \draw (5,0) coordinate (A1) -- node[above,sloped] {Region I} (1,3) coordinate (A2);
    \draw (9,0.5) coordinate (A3) -- (5,3.5) coordinate (A4);
    \draw (A1) -- (A3); \draw (A2) -- (A4);
    \fill[gray,opacity=0.3] (A1) -- (A2) -- (A4) -- (A3) -- cycle;
    \draw[dashed] (A1) -- (4,0.35) node[above] {$\theta$};
    \draw (4.25,0.55) arc (135:175:0.4);

    \draw[dashed, opacity=0.0] (5.2,0.466) coordinate (B1) -- (1.2,3.466) coordinate (B2);
    \draw[dashed, opacity=0.0] (9.2,0.966) coordinate (B3) -- (5.2,3.966) coordinate (B4);
    \draw[dashed, opacity=0.0] (B1) -- (B3); 
    \draw[dashed, opacity=0.0] (B2) -- (B4);

    \def\mypath{ (B1) [decorate, decoration={snake, amplitude=0.07cm,segment length=0.601cm}] -- (B2) -- (B4) -- (B4) -- (B3) -- (B1) }

    \fill[gray,opacity=0.2] \mypath;
    
    \draw[->] (9.2, 2) -- (9.7,2.0625) node[right] {$y$};
    \draw[->] (9.2, 2) -- (9.7, 1.625) node[right] {$x$};
    \draw[->] (9.2, 2) -- (9.4, 2.466) node[above] {$z$};
    
     \draw[->] (8.4, 3.7) -- (8.4, 3.1);
     
      \draw[-] (8.4, 3.4) node[right] {$\bm{g}$};
     
     \draw (7, 3.5) node {Region II};
     
      \draw[->] (5.4, 2.0) -- (5.7, 2.699) node[left] {$\bm{E}_0\;$};
      
       \draw[->] (6.0, 1.5) -- (6.3 , 2.199);

       \draw[->] (4.8, 2.5) -- (5.1, 3.199);
            
\end{tikzpicture}
\end{figure}
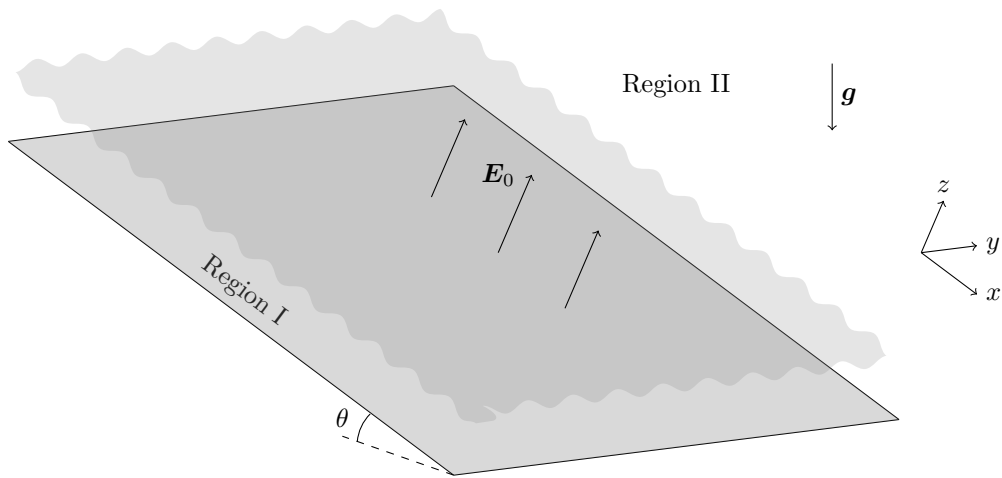

Consider a Newtonian fluid with constant density $\rho$, dynamic viscosity $\mu$, and kinematic viscosity $\nu$, flowing under gravity along a flat infinite two-dimensional substrate inclined at a non-zero angle $\theta$ to the horizontal. We use coordinates $(x,y,z)$ with $x$ directed along the angle of tilt, $y$ in the other spanwise direction, and $z$ perpendicular to the substrate, as shown for the case of an overlying film in the schematic of figure \ref{Setupdiagram}. The surface tension coefficient between the liquid and the surrounding passive medium is denoted by $\sigma$ (assumed constant), and the 
acceleration due to gravity is denoted by $\bm{g} = (g\sin\theta,0,-g\cos\theta)$. The local film thickness is denoted $h(x,y,t)$, a function of space and time, with unperturbed thickness $\ell$. The liquid film is assumed to be a perfect conductor and the surrounding medium is taken to be a perfect dielectric with permittivity $\epsilon_{\alpha}$. A voltage is set up by grounding the plate at zero potential and imposing a uniform field vertical to the plate far away, i.e. $\bm{E}\to\bm{E}_0=(0,0,E_0)$ 
as $z\to\infty$, where $E_0$ is a constant. Denoting the voltage potential by $V$, it follows that in the electrostatic limit appropriate to this study, the electric field takes the form $\bm{E} = - \bm{\nabla} V$ (this follows from Maxwell's equations that yield $\bm{\nabla}\bm{\times}\bm{E}=0$ in this limit), where $\bm{\nabla}$ is the usual three-dimensional spatial gradient operator. Since the fluid is perfectly conducting, the voltage potential is zero at the fluid interface. The liquid layer and surrounding medium are denoted by Region I and II respectively. The fluid in Region I is governed by the incompressible Navier--Stokes equations
\begin{subequations}
\begin{align} \bm{u}_t + (\bm{u} \bm{\cdot} \bm{\nabla} )\bm{u} = & -\frac{1}{\rho}{\bm{\nabla} p } + \nu {\nabla}^2 \bm{u} + \bm{g} ,\\
\bm{\nabla}\bm{\cdot}\bm{u} = & \; 0,
\end{align}
\end{subequations}
where $\bm{u}= (u,v,w)$ is the velocity field, $p$ is the pressure, and ${\nabla}^2 = \bm{\nabla} \bm{\cdot} \bm{\nabla}$. 
Note that for overlying films we have $\theta \in (0, \pi/2)$, for hanging films $\theta \in (\pi/2,\pi)$, and for vertical plates 
$\theta = \pi/2$. Since $\bm{E}=-\bm{\nabla} V$ and in addition Gauss' law states that $\bm{\nabla}\bm{\cdot} (\epsilon_{\alpha} \bm{E})=0$ 
(we assume that there are no volume charges in Region II), it follows that $V$ satisfies Laplace's equation in Region II,
\begin{equation}\label{eq:LapV}
\nabla^2 V=0,
\end{equation}
subject to the conditions
\begin{subeqnarray}\label{VBCs}
\gdef\thesubequation{\theequation \textit{a,b}}
V=0\quad\textrm{at}\quad z=h(x,y,t),\qquad \bm{\nabla} V \rightarrow - \bm{E}_0\quad\textrm{as}\quad z\rightarrow\infty.
\end{subeqnarray}
For the fluid we have no-slip conditions at the solid substrate surface, $\bm{u}|_{z=0} = 0$, the kinematic condition
\begin{equation} 
w= h_t + u h_x + v h_y \quad\textrm{at}\quad z=h(x,y,t),\label{eq:KIN}
\end{equation} 
and a balance of stresses at the interface as detailed next. Any point on the interface at time $t$ has position vector $\bm{r} = (x,y,h(x,y,t))$. The contravariant base vectors $\bm{t}_1,\bm{t}_2$ and unit normal $\bm{n}$ are defined by
\begin{equation}
\bm{t}_1 = \frac{\partial \bm{r}}{\partial x} = \begin{pmatrix} 1 \\ 0 \\ h_x \end{pmatrix}, \quad \bm{t}_2 = \frac{\partial \bm{r}}{\partial y} = \begin{pmatrix} 0 \\ 1 \\ h_y \end{pmatrix}, \quad \bm{n} = \frac{\bm{t}_1 \times \bm{t}_2}{\sqrt{K}} = \frac{1}{\sqrt{K}}\begin{pmatrix} -h_x \\ -h_y \\ 1 \end{pmatrix},
\end{equation}
where $K = 1 + h_x^2 + h_y^2$. Since the potential is constant on the interface,
\begin{subequations}\label{Vzerointerf}
\begin{equation}
\bm{\nabla} V \bm{\cdot} \bm{t}_1 = 0 \quad\Rightarrow \quad V_x + h_x V_z = 0,
\end{equation}
\begin{equation}
\bm{\nabla} V \bm{\cdot} \bm{t}_2 = 0 \quad\Rightarrow\quad V_y + h_y V_z = 0,
\end{equation}
\end{subequations}
where it is understood that all functions are evaluated at $z=h(x,y,t)$.
The stress tensors in Region I and II have components
\begin{equation}\bm{T}^{\textrm{I}}_{jk} = \mu \left( \frac{\partial u_i}{\partial x_j} + \frac{\partial u_j}{\partial x_i}\right) -  p\delta_{jk} , \quad \bm{T}^{\textrm{II}}_{jk} = \epsilon_{\alpha} \left( \frac{\partial V}{\partial x_j} \frac{\partial V}{\partial x_k} - \frac{1}{2} | \bm{\nabla}V|^2 \delta_{jk}\right) - p_{\text{atm}} \delta_{jk}, \end{equation}
respectively, where $p_{\text{atm}}$ is the atmospheric pressure in Region II and we have employed the usual subscript notation for the coordinate system and velocity components. We balance the stresses in the tangential and normal directions at the interface,
\begin{subeqnarray}\label{stressbalance}
\gdef\thesubequation{\theequation \textit{a--c}}
\left[   \left( \bm{T}^{i}\bm{n} \right) \bm{\cdot} \bm{t}_1  \right]_{\textrm{II}}^{\textrm{I}}  = 0,\quad \left[   \left( \bm{T}^{i} \bm{n} \right) \bm{\cdot} \bm{t}_2  \right]_{\textrm{II}}^{\textrm{I}}  = 0, \quad \left[ \left(\bm{T}^{i} \bm{n}\right) \bm{\cdot} \bm{n}  \right]_{\textrm{II}}^{\textrm{I}} = \sigma\kappa,
\end{subeqnarray}
where the jump notation $\left[\;\cdot\;\right]_{\textrm{II}}^{\textrm{I}}=(\;\cdot\;)_{\textrm{I}}-(\;\cdot\;)_{\textrm{II}}$ has been introduced and 
the curvature of the interface is
\begin{equation}\kappa = \frac{ (1 + h_x^2)h_{yy} - 2 h_x h_y h_{xy} + (1 + h_y^2) h_{xx} }{K^{3/2}}. \end{equation}
Using (\ref{Vzerointerf}\textit{a}), the tangential stress balance in the $\bm{t}_1$ direction (\ref{stressbalance}\textit{a}) becomes
\begin{equation}\label{tangentialstressesbal1final}(1-h_x^2)(u_z + w_x) + 2 (w_z -  u_x)h_x - (u_y + v_x)h_y - (v_z + w_y)h_x h_y = 0,\end{equation}
and similarly using (\ref{Vzerointerf}\textit{b}), the tangential stress balance in the $\bm{t}_2$ direction (\ref{stressbalance}\textit{b}) reads
\begin{equation}\label{tangentialstressesbal2final}(1 - h_y^2)(v_z + w_y) - (u_y + v_x) h_x + 2(w_z -v_y) h_y - (u_z + w_x)h_x h_y = 0.\end{equation}
The normal stress balance (\ref{stressbalance}\textit{c}) written out in full becomes
\begin{align} \label{normalstressbalfinal}
\nonumber & p_{\textrm{atm}} - p - \frac{\epsilon_{\alpha}}{K} \left[ h_x^2 \left( \frac{h_x^2}{2} + 1\right) + h_y^2 \left( \frac{h_y^2}{2} + 1\right) + h_x^2 h_y^2 + \frac{1}{2}  \right] V_z^2   \\
 \nonumber & + 2  \mu \frac{ u_xh_x^2 +  (u_y + v_x)h_xh_y +  v_yh_y^2 - (u_z + w_x)h_x - (v_z + w_y) h_y  +  w_z }{K}  \\
= \; &  \sigma \frac{ (1 + h_x^2)h_{yy} - 2 h_x h_y h_{xy} + (1 + h_y^2) h_{xx} }{K^{3/2}}.
\end{align}
The stress balances (\ref{tangentialstressesbal1final}), (\ref{tangentialstressesbal2final}) and (\ref{normalstressbalfinal}) complete the set of dimensional nonlinear interfacial conditions. The normal stress balance (\ref{normalstressbalfinal}) is the originator of the coupling between the problems in Region I and Region II. This is unique to the case of a perfectly conducting liquid film surrounded by a perfect dielectric (where one phase possesses infinite conductivity and the other zero conductivity respectively), otherwise the electric field has contributions to the tangential stresses as in the case of the Taylor--Melcher leaky dielectric model (see for example \cite{papageorgiou2004generation}).

\subsection{Exact solution and non-dimensionalisation of equations\label{ExactSolutionandDimensionlessEquationsSec}}

The exact Nusselt solution with a film of uniform thickness \cite{nusselt1,FLM:367246} can be modified to account for the
electric field as done for the one-dimensional problem in \cite{tseluiko2006wave} to give
\begin{equation}\label{basestatesdimensional} \left.
\begin{array}{c} {\displaystyle \bar{h} = \ell, \quad \bar{u} = \frac{g\sin\theta}{2\nu} (2\ell z - z^2),\quad \bar{v} = 0,\quad \bar{w} = 0,} \\[8pt]
{\displaystyle\bar{p} = p_{\textrm{atm}} - \frac{1}{2} \epsilon_{\alpha} E_0^2 - \rho g (z-\ell)\cos\theta,\quad \bar{V} = E_0(\ell - z), }
\end{array}   \right\}
\end{equation}
with bars denoting base states. The velocity profile is semi-parabolic in $z$ and the voltage potential is linear in $z$ as expected. We will non-dimensionalise velocities with the base velocity at the free surface,
\begin{equation}\label{baseU0rescale}
U_0 = \bar{u}|_{z = \ell} = \frac{g \ell^2 \sin\theta}{2\nu}.
\end{equation}
More specifically we write
\begin{equation} 
\left. \begin{array}{c} {\displaystyle x^{*} = \frac{1}{\ell} x, \quad y^{*} = \frac{1}{\ell} y, \quad z^{*} = \frac{1}{\ell} z, \quad \bm{u}^{*} = \frac{1}{U_0} \bm{u},} \\[8pt]
{\displaystyle t^{*} = \frac{U_0}{\ell} t , \quad p^{*} = \frac{1}{\rho U_0^2} p, \quad V^{*} = \frac{1}{E_0\ell} V, \quad h^{*} = \frac{1}{\ell} h,}\end{array}\right\}  
\end{equation}
substitute into the equations and boundary conditions, and drop the stars. In Region I, the Navier--Stokes equations transform to
\begin{subequations}
\begin{align} \bm{u}_t + (\bm{u} \bm{\cdot} \bm{\nabla} )\bm{u} = & - \bm{\nabla} p  + \frac{1}{{\Rey}} {\nabla}^2 \bm{u} +  \frac{2}{{\Rey}} \bm{f} ,\\
\bm{\nabla}\bm{\cdot}\bm{u} = & \; 0 ,
\end{align}
\end{subequations}
where $\bm{f} = (1,0,-\cot\theta)$. Laplace's equation in Region II and the no-slip and impermeability conditions are unchanged, while 
the far field condition for $V$ (\ref{VBCs}\textit{b}) becomes $\bm{\nabla} V \rightarrow (0,0,-1)$ as $z\rightarrow \infty$. 
At the interface, the zero voltage potential condition (\ref{VBCs}\textit{a}), kinematic condition \eqref{eq:KIN}, and the tangential stress relations
\eqref{tangentialstressesbal1final} and \eqref{tangentialstressesbal2final} are all unchanged, whereas the normal stress relation (\ref{normalstressbalfinal}) transforms to (all variables evaluated at $z=h$)
\begin{align}
\nonumber & \frac{1}{2} {\Rey} \left( \bar{p}_{\textrm{atm}} - p\right) - \frac{{\Weber}}{K} \left[ h_x^2 \left( \frac{h_x^2}{2} + 1\right) + h_y^2 \left( \frac{h_y^2}{2} + 1\right) + h_x^2 h_y^2 + \frac{1}{2}  \right] V_z^2   \\
\nonumber & +  \frac{ u_xh_x^2 +  (u_y + v_x)h_xh_y +  v_yh_y^2 - (u_z + w_x)h_x - (v_z + w_y) h_y  +  w_z }{K}  \\
= \; &  \frac{1}{2{\Capil}} \frac{ (1 + h_x^2)h_{yy} - 2 h_x h_y h_{xy} + (1 + h_y^2) h_{xx} }{K^{3/2}},
\end{align}
where $\bar{p}_{\textrm{atm}} = p_{\textrm{atm}}/(\rho U_0^2)$ is the non-dimensional constant pressure in Region II. The other dimensionless parameters are
\begin{equation}\label{dimensionlessparameters} {\Rey} = \frac{U_0 \ell}{\nu} = \frac{g \ell^3 \sin\theta}{2\nu^2},\qquad {\Weber} = \frac{\epsilon_{\alpha} E_0^2}{\rho g \ell \sin\theta}, \qquad {\Capil} = \frac{U_0 \mu}{\sigma} = \frac{ \rho g \ell^2 \sin\theta }{2\sigma},\end{equation}
where ${\Rey}$ is the Reynolds number measuring the ratio of inertial to viscous forces, 
${\Weber}$ is the electric Weber number measuring the ratio of electrical to fluid pressures, 
and ${\Capil}$ is the capillary number measuring the ratio of surface tension to viscous forces. 

Writing the solution as
\begin{equation} \left. \begin{array}{c} {\displaystyle \label{disturbancechangeofvars}  u = z(2 - z) + \tilde{u} ,\qquad v = \tilde{v},\qquad w = \tilde{w},} \\[8pt]
{\displaystyle p =  \bar{p}_{\textrm{atm}} - \frac{{\Weber} }{ {\Rey} } - \frac{  2 }{ {\Rey}} ( z -  1)\cot\theta + \tilde{p} ,\qquad V = 1 - z + \tilde{V},} \end{array} \right\} \end{equation}
where tilde quantities are of order one (these are perturbations of the non-dimensional base states), substituting
into the non-dimensional Navier--Stokes equations in Region I and dropping the tildes yields the following equations for the perturbations,
\begin{subequations}
\begin{align} \bm{u}_t + ((\bm{u} + \bm{\phi})\bm{\cdot} \bm{\nabla} ) (\bm{u}+\bm{\phi}) = & - \bm{\nabla} p + \frac{1}{{\Rey}} {\nabla}^2 \bm{u}, \\
\bm{\nabla} \bm{\cdot} \bm{u} = & \; 0,
\end{align}
\end{subequations}
where $\bm{\phi} = (z(2-z),0,0)$. Laplace's equation in Region II and the no-slip and impermeability conditions are unchanged under this change of variables, while the far field condition becomes $\bm{\nabla}V \rightarrow \bm{0}$ as $z\rightarrow \infty$. Additionally, we substitute (\ref{disturbancechangeofvars}) into the equations at the interface, and drop the tildes. We obtain the following equations for the 
nonlinear perturbations at the interface (all variables are evaluated at $z=h$),
\begin{subequations}
\begin{equation}V = h-1,\end{equation}
\begin{equation}\label{kinematicperturb1}w = h_t + (h(2-h) + u)h_x + vh_y,\end{equation}
\begin{equation}(1-h_x^2)(2(1 - h) + u_z + w_x) + 2 (w_z -  u_x)h_x - (u_y + v_x)h_y - (v_z + w_y)h_x h_y = 0,\end{equation}
\begin{equation}(1 - h_y^2)(v_z + w_y) - (u_y + v_x) h_x + 2 (w_z -v_y) h_y - (2(1-h) + u_z + w_x)h_x h_y = 0,\end{equation}
\begin{align}
\nonumber & \frac{1 }{ 2 }{\Weber}  + ( h -  1)\cot\theta -  \frac{1}{2} {\Rey}\; p  \\
\nonumber & - \frac{{\Weber}}{K} \left[ h_x^2 \left( \frac{h_x^2}{2} + 1\right) + h_y^2 \left( \frac{h_y^2}{2} + 1\right) + h_x^2 h_y^2 + \frac{1}{2}  \right] (1 - V_z)^2   \\
\nonumber & +  \frac{ u_xh_x^2 +  (u_y + v_x)h_xh_y +  v_yh_y^2 - (2( 1 - h) + u_z + w_x)h_x - (v_z + w_y) h_y  +  w_z }{K}  \\
= \; &  \frac{1}{2{\Capil}} \frac{ (1 + h_x^2)h_{yy} - 2 h_x h_y h_{xy} + (1 + h_y^2) h_{xx} }{K^{3/2}}.
\end{align}
\end{subequations}
The system remains nonlinear and intractable analytically; in what follows we make progress by
considering nonlinear long wave disturbances at the interface, that is the typical lengths in the $x$ and $y$ directions are large compared to the film thickness.

\section{Fully nonlinear long-wave evolution equations\label{ALong-WaveEvolutionEquationSec}}

We assume that the typical interfacial deformation wavelengths $\lambda$ are large compared to the unperturbed thickness $\ell$, set $\delta = \ell /\lambda \ll 1$, and introduce the following change of variables in Region I,
\begin{equation}\label{longwavereg1changeofvars} 
x = \frac{1}{\delta} \hat{x},\qquad y = \frac{1}{\delta} \hat{y}, \qquad t = \frac{1}{\delta} \hat{t},\qquad w = \delta \hat{w}. 
\end{equation}
For brevity we omit the transformed Navier--Stokes equations. The no-slip and impermeability conditions are unchanged. Substitution of (\ref{longwavereg1changeofvars}) into the interfacial conditions and dropping hats keeps the kinematic condition unchanged, while the stress balances read (all variables evaluated at $z=h$):
\begin{subequations}
\begin{equation}\label{tangentialstressdir1}(1-\delta^2 h_x^2)(2(1 - h) + u_z + \delta^2 w_x) + 2 \delta^2 ( w_z - u_x)h_x - \delta^2 (u_y + v_x)h_y - \delta^2 (v_z + \delta^2 w_y)h_x h_y = 0,\end{equation}
\begin{equation}\label{tangentialstressdir2}(1 - \delta^2 h_y^2)(v_z + \delta^2 w_y) - \delta^2 (u_y + v_x) h_x + 2\delta^2 (w_z -v_y) h_y - \delta^2 (2(1-h) + u_z + \delta^2 w_x)h_x h_y = 0,\end{equation}
\begin{align}\label{scalednormalinterface}
\nonumber & \frac{1 }{ 2 }{\Weber}  + ( h -  1)\cot\theta -  \frac{1}{2} {\Rey}\;  p  \\
\nonumber & - \frac{{\Weber}}{K_{\delta}} \left[ \delta^2 h_x^2 \left( \delta^2\frac{h_x^2}{2} + 1\right) + \delta^2h_y^2 \left( \delta^2\frac{h_y^2}{2} + 1\right) + \delta^4 h_x^2 h_y^2 + \frac{1}{2}  \right] (1 - V_z)^2   \\
\nonumber & + \frac{1}{K_{\delta}} \left[ \delta^3 u_xh_x^2 +  \delta^3(u_y + v_x)h_xh_y + \delta^3 v_yh_y^2 \right. \\
\nonumber & \left. \qquad\qquad\qquad - \delta (2( 1 - h) + u_z + \delta^2 w_x)h_x - \delta (v_z + \delta^2 w_y) h_y  +  \delta w_z \right]  \\
= \; &  \frac{1}{2{\Capil}} \frac{ \delta^2 (1 +  \delta^2 h_x^2)h_{yy} - 2  \delta^4 h_x h_y h_{xy} +  \delta^2(1 +  \delta^2 h_y^2) h_{xx} }{K_{\delta}^{3/2}},
\end{align}
\end{subequations}
where $K_{\delta} = \delta^2 h_x^2 + \delta^2 h_y^2 + 1$. The normal stress balance 
(\ref{scalednormalinterface}) contains a non-local contribution since $V$ satisfies Laplace's equation in Region II. 
This non-local contribution is calculated by introducing the following variables in Region II,
\begin{equation} 
x = \frac{1}{\delta} \hat{x},\qquad y = \frac{1}{\delta} \hat{y}, \qquad z = \frac{1}{\delta} \hat{z}.
\end{equation}
The problem for the perturbation potential becomes
\begin{equation} \left. \begin{array}{c} {\displaystyle {\nabla}^2V = 0,}\\[8pt]
{\displaystyle \bm{\nabla}V \rightarrow \bm{0} \,\, \textrm{ as }\,\, z\rightarrow \infty,}\\[8pt]
{\displaystyle V|_{z = \delta h} = h - 1,}\end{array} \right\}
\end{equation}
from which we need to obtain a leading order approximation of $V_z|_{z=\delta h }$ to use in the normal stress balance
equation \eqref{scalednormalinterface}. Introducing the asymptotic expansions
\begin{equation} \label{expansionsinregion2}  h = h_0 + \delta h_1 + \delta^2 h_2 + \ldots,\qquad  V = V_0 + \delta V_1 + \delta^2 V_2 + \ldots.\end{equation}
and noting that $V|_{z=\delta h} = V_0|_{z=0} + O(\delta)$, $V_z|_{z=\delta h} = (V_0)_z|_{z=0} + O(\delta)$, yields the
leading order problem
\begin{equation} \left. \begin{array}{c} {\displaystyle {\nabla}^2V_0 = 0, }\\[8pt]
 \label{problemforV0DtoN}{\displaystyle \bm{\nabla}V_0 \rightarrow \bm{0} \,\,\textrm{ as }\,\, z\rightarrow \infty,}\\[8pt]
{\displaystyle V_0 |_{z = 0} = h_0 - 1.}
\end{array} \right\}
\end{equation}
From the Poisson formula, $(V_0)_z |_{z=0} = \mathcal{R}(h_0 - 1)$, where $\mathcal{R}$ is the Dirichlet-to-Neumann map defined by
\begin{equation}\label{operatorRdefinition}\mathcal{R}(f) = \frac{1}{2\pi}  \int_{\mathbb{R}^2} \frac{f(\bm{x}')}{ |\bm{x} - \bm{x}'|^3} \; \mathrm{d}\bm{x}',\end{equation}
where the integral is understood in a distributional sense and $\bm{x} = (x,y)$, $\bm{x}' = (x',y')$. This integral representation is not particularly useful. The Fourier symbol of the operator $\mathcal{R}$ can be obtained from the original Laplace problem by taking Fourier transforms 
in $x$ and $y$, and considering the resulting differential equation to obtain 
$\widehat{\mathcal{R}}= - |\bm{\xi}| = - \sqrt{ \xi_1^2 + \xi_2^2 }$ for wavenumber vector $\bm{\xi} = (\xi_1,\xi_2)$. We expand on the important properties of this non-local operator in Appendix \ref{PropertiesofRappendix}. It follows that
\begin{equation}\label{order1efieldeffect}
V_z|_{z=\delta h} = E^{\textrm{II}} := \mathcal{R}(h_0 - 1) + O(\delta).
\end{equation}
In terms of the scaled variables (\ref{longwavereg1changeofvars}) in Region I, 
this non-local contribution transforms to $\delta E^{\textrm{II}}$. It follows from (\ref{scalednormalinterface}) 
that in order to retain the effects of surface tension and the electric field in the leading order dynamics, we must take the scalings
\begin{equation}{\Capil} = \delta^2 \overline{{\Capil}}, \quad {\Weber} = \frac{\overline{{\Weber}}}{\delta},\end{equation}
where $ \overline{{\Capil}}$ and $\overline{{\Weber}}$ are $O(1)$ quantities. We also assume that the Reynolds number ${\Rey}$ is an $O(1)$ quantity.

Turning to the fluid dynamics in Region I, we introduce the following asymptotic expansions
\begin{equation} 
\left. \begin{array}{c} {\displaystyle \label{expansionsinregion1} \;  u = u_0 + \delta u_1 + \delta^2 u_2 + \ldots ,\qquad  v = v_0 + \delta v_1 + \delta^2 v_2 + \ldots ,} \\ 
 w = w_0 + \delta w_1 + \delta^2 w_2 + \ldots ,\qquad  p = p_0 + \delta p_1 + \delta^2 p_2 + \ldots. \end{array} \right\}
 \end{equation}
To leading order, the kinematic equation \eqref{kinematicperturb1} becomes 
\begin{equation}\label{kinematich0unsimplified} 
(h_0)_t + \left[h_0(2-h_0) + u_0\right](h_0)_x + v_0(h_0)_y - w_0 = 0\quad\textrm{at}\quad z=h_0(x,y,t).
\end{equation}
The leading order terms from the spanwise momentum equations are
\begin{equation}
(u_0)_{zz} = 0,\qquad (v_0)_{zz} = 0.
\end{equation}
These can be integrated to obtain
\begin{equation} 
(u_0)_z = 2(h_0 - 1), \qquad (v_0)_z = 0,
\end{equation}
where we have used the leading order terms of the tangential stress balances 
(\ref{tangentialstressdir1}) and (\ref{tangentialstressdir2}),
\begin{equation}
2(1-h_0) + (u_0)_z|_{z=h_0} = 0, \quad (v_0)_z|_{z=h_0} = 0. 
\end{equation}
One more integration, use of no-slip and the leading order continuity equation provides
the leading order flow field
\begin{equation}\label{u0andv0asymp} 
u_0 = 2(h_0 - 1)z, \qquad v_0 = 0,\qquad w_0 = -z^2(h_0)_x.
\end{equation}
Substituting  (\ref{u0andv0asymp}) into the leading order kinematic equation (\ref{kinematich0unsimplified}) yields
\begin{equation}\label{h0kinematic}
(h_0)_t + 2h_0^2(h_0)_x = 0.
\end{equation}
We need to regularise this equation by adding higher order terms since its solutions encounter infinite slope singularities at finite times and 
the long-wave expansion breaks down. Note that at leading order, the $z$-momentum equation implies that $p_0$ is independent of $z$, 
so to leading order the normal stress balance (\ref{scalednormalinterface}) gives
\begin{equation} 
p_0 = \frac{2}{{\Rey}} \left[ (h_0-1)\cot\theta + \overline{{\Weber}} \mathcal{R}(h_0-1) - \frac{1}{2\overline{{\Capil}}} ((h_0)_{xx} + (h_0)_{yy}) \right].
\end{equation}
We proceed as before but now collect $O(\delta)$ terms in the governing equations and boundary conditions. 
The second-order contribution to the kinematic condition (\ref{kinematicperturb1}) is found to be (note that Taylor expansions about
$z=h_0$ are used)
\begin{equation} \label{kinematic1storder} 
(h_1)_t  +  u_1|_{z=h_0} (h_0)_x + 2 h_0  h_1 (h_0)_x + h_0^2 (h_1)_x + v_1|_{z =h_0} (h_0)_y - w_1 |_{z = h_0} = 0.
\end{equation}
As above, the first order velocities $u_1$, $v_1$ and $w_1$ can be found analytically by integration of
the second-order momentum equations (and using tangential stresses, no slip and
the continuity equation). For completeness, these are
\begin{subequations}
\begin{equation}\label{u1asymp} u_1 = \frac{1}{2} {\Rey} z^2 (p_0)_x - {\Rey}h_0 z (p_0)_x -  \frac{2}{3} {\Rey} z^3 h_0^2 (h_0)_x +  \frac{1}{6} {\Rey} z^4 h_0 (h_0)_x +  \frac{4}{3} {\Rey} z h_0^4 (h_0)_x + 2 z h_1,\end{equation}
\end{subequations}
\begin{subeqnarray}\addtocounter{equation}{-1}
\gdef\thesubequation{\theequation \textit{b,c}}
v_1 = \frac{1}{2} {\Rey} z^2 (p_0)_y  - {\Rey} h_0 z (p_0)_y,\quad w_1=-\int_0^z \left\{(u_1)_x+(v_1)_y\right\} \mathrm{d}z'. 
\end{subeqnarray}
Substituting these into 
(\ref{kinematic1storder}) gives
\begin{equation}\label{h1kinematic}
(h_1)_t  + \left[ \frac{8}{15} {\Rey} h_0^6 (h_0)_x + 2 h_0^2 h_1 -  \frac{1}{3} {\Rey} h_0^3 (p_0)_x \right]_{x} + \left[ - \frac{1}{3} {\Rey} h_0^3 (p_0)_y \right]_y = 0.
\end{equation}
A regularised Benney equation for $H = h_0 + \delta h_1$, correct to $O(\delta^2)$, is found by adding $\delta$ times equation 
(\ref{h1kinematic}) to (\ref{h0kinematic}), this is
\begin{align} \nonumber H_t + \left[ \frac{2}{3} H^3 + \frac{8{\Rey}}{15} \delta  H^6 H_x -  \frac{2}{3} \delta H^3 \left[ H_x\cot\theta + \overline{{\Weber}} \mathcal{R}(H-1)_x - \frac{1}{2\overline{{\Capil}}} (H_{xxx} + H_{xyy}) \right] \right]_{x} & \\ + \label{benneyqequation1} \left[ -  \frac{2}{3} \delta H^3 \left[ H_y\cot\theta + \overline{{\Weber}} \mathcal{R}(H-1)_y - \frac{1}{2\overline{{\Capil}}} (H_{xxy} + H_{yyy}) \right] \right]_y = 0. & \end{align}
Tseluiko and Papageorgiou \cite{tseluiko2006wave} noted for the one-dimensional analogue of (\ref{benneyqequation1}) that solutions may not exist for all time, for some parameters, and finite-time blow-ups are observed in numerical simulations. Due to such global existence difficulties
we proceed by studying the weakly nonlinear evolution of a sufficiently small perturbation to the uniform state. 
The above procedure was also carried out to the next order in $\delta$ to calculate a Benney equation which is accurate to $O(\delta^3)$; 
this is required to retain dispersive effects. This equation is currently under investigation and findings will be reported in future work.

\section{A multidimensional non-local Kuramoto--Sivashinsky equation\label{ANon-localKuramoto-SivashinskyEquationSec}}

\subsection{Weakly nonlinear evolution} 

We substitute $H = 1 + \alpha(\delta) \eta$ into equation (\ref{benneyqequation1}) where $\alpha(\delta) = o(\delta^{1/2})$ is a positive scaling parameter and $\eta = O(1)$ - see \cite{tseluiko2006wave} for the one-dimensional equation. 
Correct to $O(\delta)$, the resulting equation is
\begin{equation} \eta_t +  2 \eta_x + 4 \alpha \eta \eta_x + \frac{8 {\Rey}}{15} \delta \eta_{xx}  -  \frac{2}{3} \delta  \eta_{xx} \cot\theta -  \frac{2}{3} \delta  \eta_{yy} \cot\theta - \frac{2\overline{{\Weber}}}{3}   \delta  \Delta \mathcal{R}(\eta) +  \frac{1}{3\overline{{\Capil}}} \delta \Delta^2 \eta   = 0 ,\end{equation}
where $\Delta\equiv \partial_x^2+\partial_y^2$ is the usual two-dimensional Laplace operator.
Conservation of mass implies that 
$\eta$ has zero spatial mean. Rescaling with
\begin{equation}\label{rescale100} 
\overline{t} =  4\alpha\delta t, \qquad \overline{x} = x -2t , \qquad \overline{\eta} =  \frac{\eta}{\delta}, 
\end{equation}
and dropping bars gives
\begin{equation} \label{illposed2delectric} 
\eta_t + \eta \eta_x + (\beta^* - \kappa) \eta_{xx}  -  \kappa  \eta_{yy}  - \gamma^*  \Delta \mathcal{R}(\eta) + \mu \Delta^2 \eta  = 0,
\end{equation}
where
\begin{equation}\label{rescaledparameters}
\beta^* = \frac{2{\Rey}}{15\alpha} ,\qquad \kappa = \frac{1}{6\alpha}\cot\theta, \qquad \gamma^* = \frac{\overline{{\Weber}}}{6\alpha}, \qquad  \mu = \frac{1}{12\alpha\overline{{\Capil}}}.
\end{equation}
It is clear from our previous rescalings that $\beta^*, \mu > 0$, $\gamma^* \geq 0$, and that $\kappa > 0$, $\kappa = 0$, or $\kappa < 0$ depending on whether the film is overlying, vertical, or hanging respectively. Note that if the electric field is removed and we also consider a vertical substrate, setting $\gamma^* = 0$ and $\kappa = 0$ in (\ref{illposed2delectric}), then, after rescaling, the two-dimensional Kuramoto--Sivashinsky equation obtained by Nepomnyashchy \cite{Nepo1, Nepo2} is recovered,
\begin{equation}\eta_t + \eta\eta_x + \eta_{xx} + \Delta^2 \eta = 0. \end{equation}
The operator corresponding to the linear part of (\ref{illposed2delectric}),
\begin{equation} \mathcal{L} = (\beta^* - \kappa) \partial_{xx}  -  \kappa  \partial_{yy }- \gamma^*  \Delta \mathcal{R}(\bm{\cdot}) + \mu \Delta^2\end{equation}
has Fourier symbol
\begin{equation}\label{illposed2delectricsymb} 
\widehat{\mathcal{L}}(\bm{\xi}) =  - (  \beta^* - \kappa ) \xi_1^2 + \kappa \xi_2^2 - \gamma^* ( \xi_1^2 + \xi_2^2)^{3/2} + \mu (\xi_1^2 + \xi_2^2)^2, 
\end{equation}
for wavenumber vector $\bm{\xi} = (\xi_1,\xi_2)$. If we consider hanging films with $\kappa < 0$, then it is clear that there are linearly unstable $y$-modes (Fourier modes which are purely transverse). Even for overlying films with non-zero values of $\gamma^*$ and sufficiently small values of the product $\kappa \mu $, a band of low $y$-modes are linearly unstable due to the non-local term corresponding to the electric field. Due to the form of the nonlinearity in (\ref{illposed2delectric}), there is no energy transfer between $y$-modes. 
Thus if a $y$-mode is linearly unstable, then it will grow exponentially without bound and the problem is
ill-posed in this sense. There is no control over these transverse instabilities, and the weakly nonlinear 
analysis cannot be modified to overcome this issue. Since the case of $\gamma^*=0$ is not of particular interest, we are forced to restrict to overlying films with $\kappa > 0$ by taking $\theta \in (0, \pi/2)$. Then we rescale (\ref{illposed2delectric}) with
\begin{equation}\label{rescale200} 
\overline{t} =  \frac{\kappa^2}{\mu} t, \quad \overline{x} =  \frac{\kappa^{1/2}}{\mu^{1/2}}x , \quad \overline{y} =  \frac{\kappa^{1/2}}{\mu^{1/2}}y , \quad \overline{\eta} = \frac{\mu^{1/2}}{\kappa^{3/2}} \eta,
\end{equation}
and once again drop the bars, to obtain the following canonical equation for overlying electrified films,
\begin{equation} \label{illposed2delectric2} 
\eta_t + \eta \eta_x + (\beta -1) \eta_{xx}  -  \eta_{yy}  - \gamma  \Delta \mathcal{R}(\eta) + \Delta^2 \eta  = 0.
\end{equation}
The parameters $\beta > 0$, $\gamma \geq 0$ are defined by
\begin{equation}\label{rescaledparameters2}\beta = \frac{\beta^*}{\kappa} , \quad \gamma = \frac{\gamma^*}{\kappa^{1/2}\mu^{1/2}}.\end{equation}
To prevent unbounded growth of solutions, (\ref{illposed2delectric2}) is studied under certain restrictions on $\gamma$ and domain choice
(both $\gamma$ and the domain dimensions affect the unstable spectrum as we see below). 
We proceed with $Q$-periodic domains (where $Q=[0,L_1]\times [0,L_2]$) for which there are no unstable 
$y$-modes for the choices of $L_1$ and $L_2$, leaving us only with a restriction on $\gamma$. 
In what follows we perform a thorough linear stability analysis expanding on the above discussion to determine this condition. 
Clearly this issue of transverse instability caused by the electric field or hanging arrangements is unique to the full three-dimensional problem, since all the instabilities are controlled by the energy transfer due to the nonlinear term for the two-dimensional analogue where the interface dynamics are governed by (\ref{KSE1dE1}).

\begin{figure}
\caption{Linear stability regions for $\gamma = 2$ and a range of Reynolds numbers. The number of unstable modes within regions in the $\nu_1$--$\nu_2$ plane is displayed, where we have only counted the pairs or quartets of modes as one. The diagonal lines correspond to $\nu_1=\nu_2$ along which we perform numerical simulations.} \label{gamma=2_stability}
\begin{subfigure}{2.6in}
\caption{$\beta = 0.01$}
\includegraphics[width=2.6in, trim=0cm 0cm 0cm 1cm, clip=true]{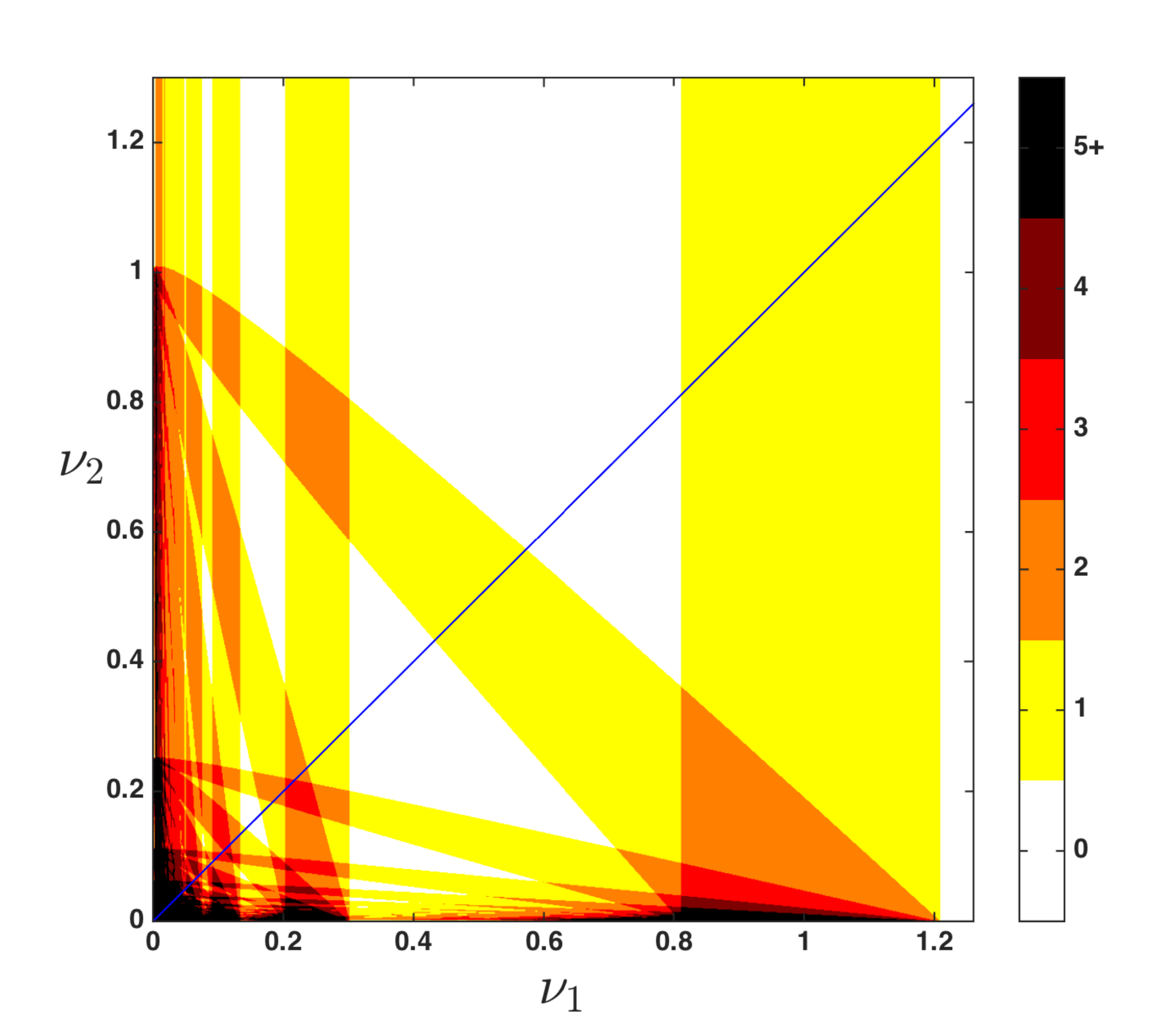}
\end{subfigure}
\begin{subfigure}{2.6in}
\caption{$\beta = 0.5$}
\includegraphics[width=2.6in, trim=0cm 0cm 0cm 1cm, clip=true]{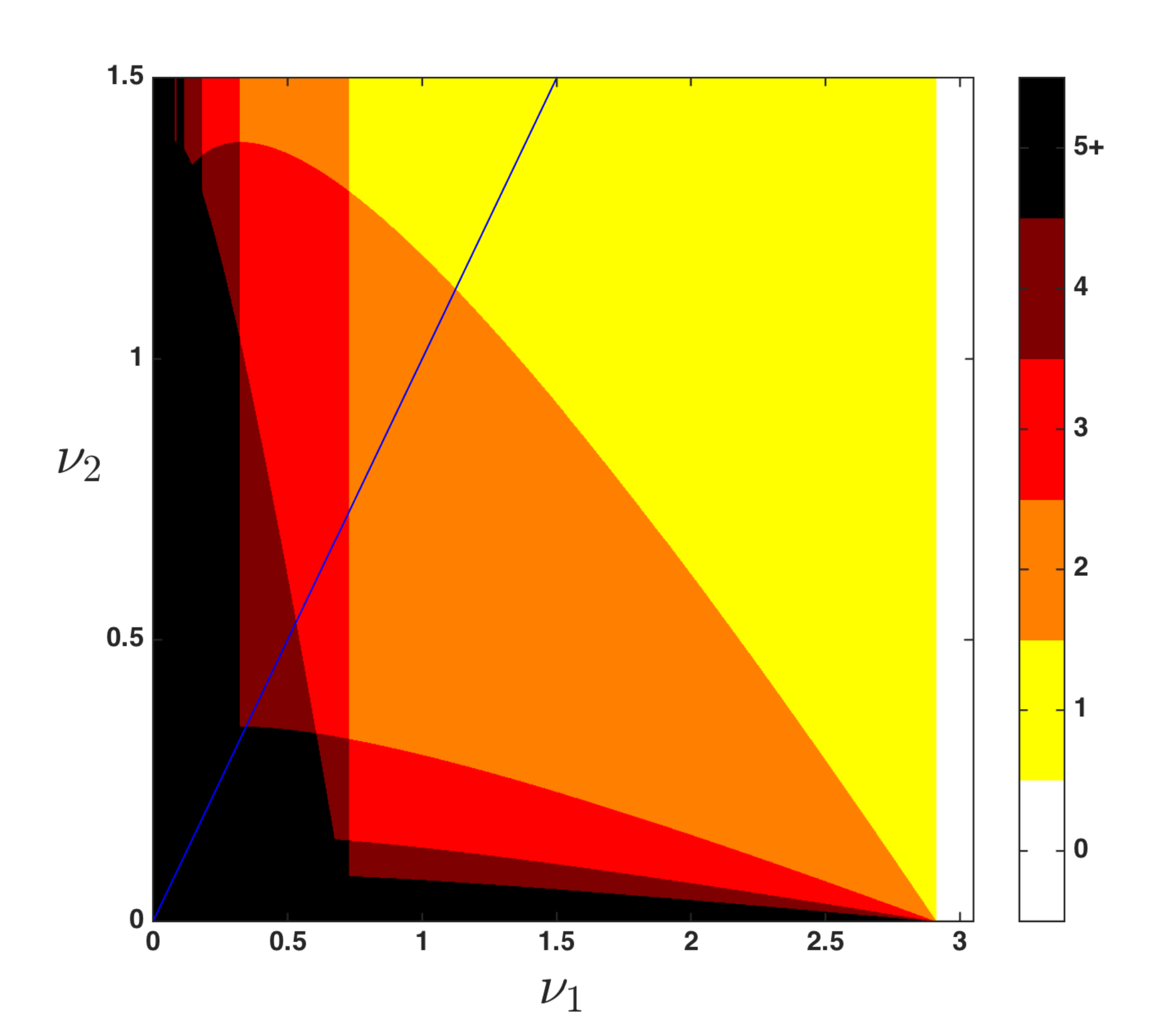}
\end{subfigure}
\vskip+0.5cm
\begin{subfigure}{2.6in}
\caption{$\beta = 1$}
\includegraphics[width=2.6in, trim=0cm 0cm 0cm 1cm, clip=true]{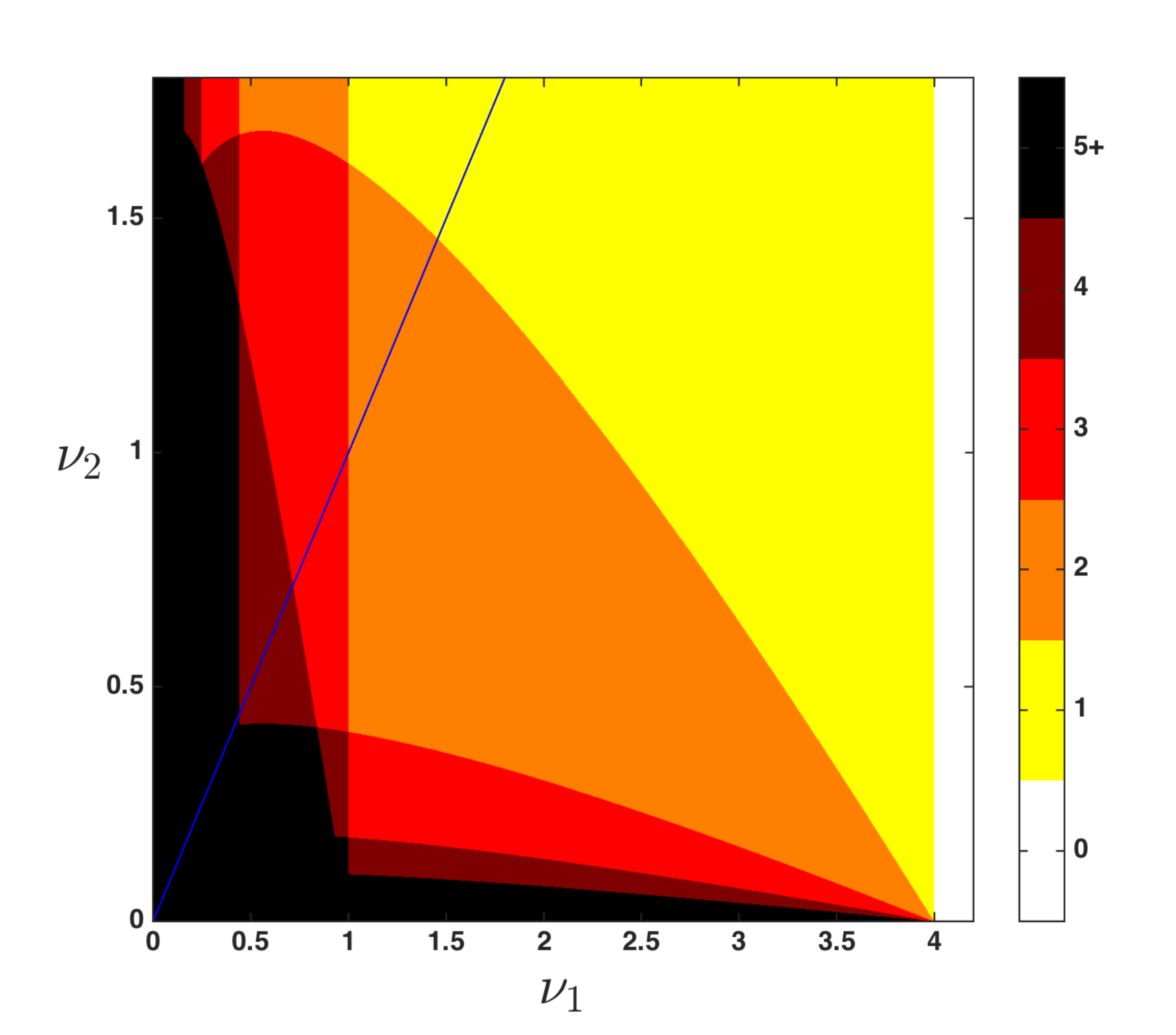}
\end{subfigure}
\begin{subfigure}{2.6in}
\caption{$\beta = 2$}
\includegraphics[width=2.6in, trim=0cm 0cm 0cm 1cm, clip=true]{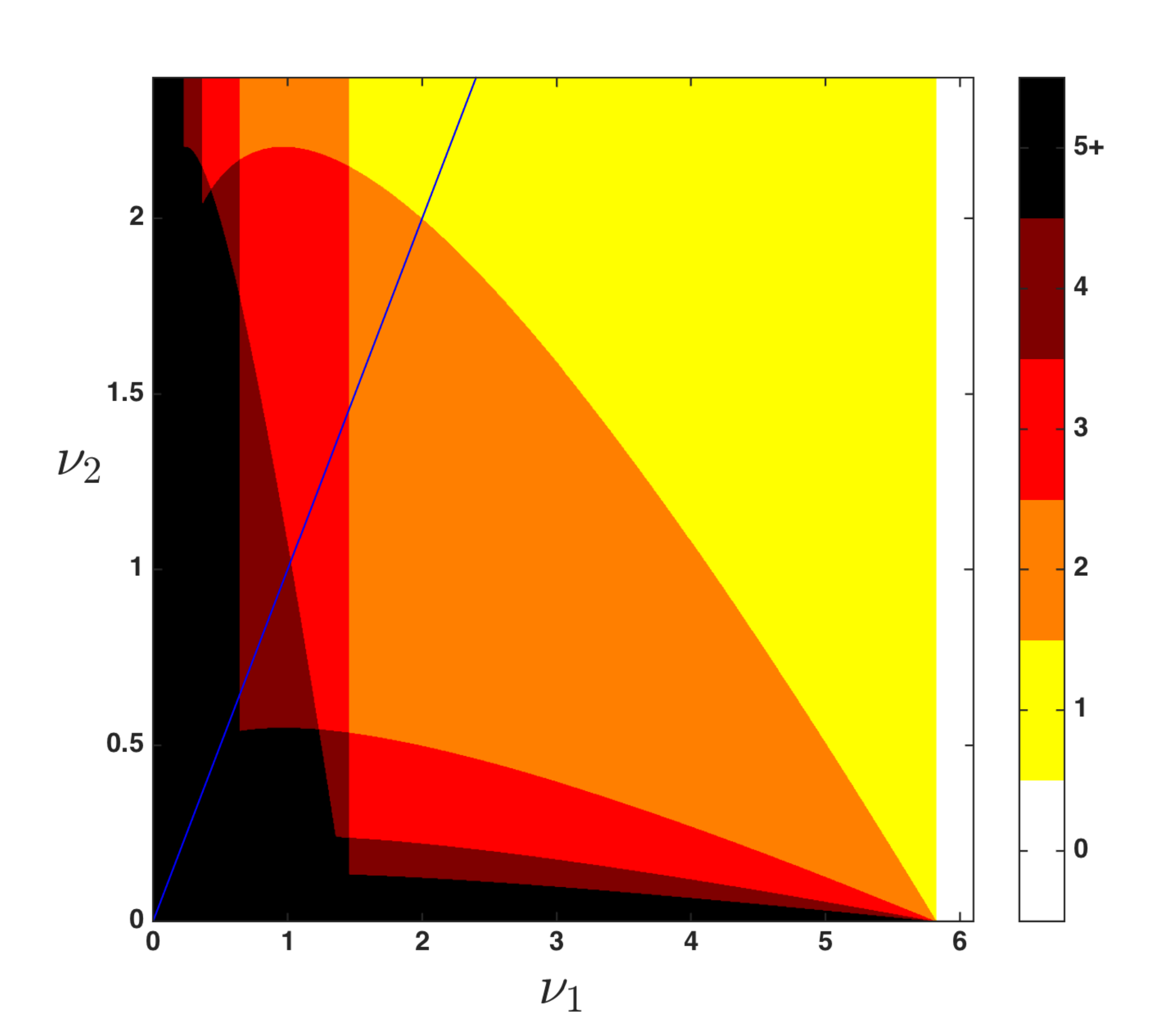}
\end{subfigure}
\end{figure}

\subsection{Linear stability analysis}

We linearise (\ref{illposed2delectric2}) about $\eta = 0$ to find
\begin{equation} \label{2Dlinear} 
\eta_t + (\beta - 1) \eta_{xx} - \eta_{yy} - \gamma  \Delta \mathcal{R}(\eta) + \Delta^2 \eta   = 0, 
\end{equation}
and look for solutions of the form
\begin{equation}\label{Fourierseriescharacter}\eta(\bm x,t) = \sum_{\bm{k}\in\mathbb{Z}^2} A_{\bm{k}}e^{i \bm{\tilde{k}}\bm{\cdot} \bm{x}+st},\end{equation} 
where $s\left(\bm{\tilde{k}}\right)$ is the growth rate, $A_{\bm{k}}$ are constants, and $\bm{\tilde{k}}$ is defined by
\begin{equation}\label{ktildedefn}\tilde{k}_j = \frac{2\pi }{L_j} k_j,\quad j=1,2.\end{equation}
Using the properties of the operator $\mathcal{R}$, the dispersion relation follows readily,
\begin{equation}\label{illposed2delectricsymb2} 
s\left(\tilde{k}_1,\tilde{k}_2\right) =  (  \beta - 1 ) \tilde{k}_1^2 -  \tilde{k}_2^2 + \gamma \left( \tilde{k}_1^2 + \tilde{k}_2^2\right)^{3/2} -  
\left(\tilde{k}_1^2 + \tilde{k}_2^2\right)^2. 
\end{equation}
Letting $\nu_1 = (2\pi/L_1)^2$ and $\nu_2 = (2\pi/L_2)^2$ casts \eqref{illposed2delectricsymb2} into
\begin{equation} \label{sk}
s(k_1,{k}_2) = (\beta - 1) \nu_1k_1^2-\nu_2 k_2^2+\gamma ( \nu_1k_1^2 + \nu_2k_2^2)^{3/2} -  (\nu_1k_1^2 + \nu_2k_2^2)^2, \end{equation}
where $k_1$ and $k_2$ are integers. Given domain dimensions $L_1$, $L_2$, fixes the values of $\nu_1, \nu_2$ and instability is
found when $s(k_1,k_2)>0$ (note that $s$ is real). Neutral stability curves for a given mode $(k_1,k_2)$ in the $\nu_1$--$\nu_2$ plane 
follow by setting  $s(k_1,k_2)=0$ in \eqref{sk} above. Note that the neutral stability curve for the $(k_1,k_2)$-mode is the same as the neutral stability curve for the $(|k_1|,|k_2|)$-mode, so we refer to the latter for simplicity. It is straightforward to calculate the neutral stability curves for $x$-modes or $y$-modes 
(purely streamwise, $k_2=0$, or purely transverse waves,
$k_1=0$). For the $(k_1,0)$-mode these are straight lines defined by
\begin{equation}
\nu_1^{\pm} = \frac{\gamma^2+ 2(\beta - 1) \pm \gamma \sqrt{\gamma^2  + 4  (\beta - 1)} }{2 k_1^2}
\end{equation}
for parameters such that the right hand side is real. Then, if $\gamma^2  + 4 (\beta - 1) \leq 0$, these modes are always linearly stable. 
If $ \gamma^2  + 4  (\beta - 1)>0 $, then the region of linear instability is a strip defined by $\nu_1^{-}  < \nu_1 < \nu_1^{+}$; there is a strip of linear stability for these modes at small $\nu_1$ unless $\nu_1^{-} \leq 0$. 
Similarly for the $(0,k_2)$-mode, equation \eqref{sk} gives the straight line neutral curves defined by
$2 k_2 \left(\nu_2^{\pm}\right)^{1/2}=\gamma\pm\sqrt{\gamma^2-4}$, and it follows that
we have linear stability for $\gamma  \leq 2$, while for  $\gamma>2$ there is a strip of linear instability in the $\nu_2-$interval
\begin{equation}
\frac{\gamma^2 -2- \gamma \sqrt{\gamma^2  - 4} }{2 k_2^2} < \nu_2 < 
\frac{\gamma^2-2 + \gamma \sqrt{\gamma^2 -4}}{2 k_2^2}.
\end{equation}
Hence $\gamma \leq 2$ is precisely the condition we need to impose in order to study (\ref{illposed2delectric2}) for any domain dimensions; the condition ensures that the $y$-modes are damped for $\gamma < 2$ or 
neutral at distinct values of $L_2$ for $\gamma = 2$. This restriction on $\gamma$ translates back to the condition
\begin{equation}\overline{{\Weber}} \leq \left( \frac{2 \cot\theta }{\overline{{\Capil}}} \right)^{1/2}.\end{equation}
It is important to note that this does not mean that the mixed Fourier modes are also linearly stable. 
Finding the neutral stability curves for these is a computational problem and for particular values of the parameters $\beta$ and $\gamma$, the regions of stability in the $\nu_1$--$\nu_2$ plane are quite complicated. 
Recall that $\beta = 1$ corresponds to taking the critical Reynolds number for the flow,
${\Rey}_{c} = 5 \cot\theta/4$, with $\beta<1$ ($\beta>1$) being subcritical (supercritical).
For the subcritical case we will show numerical simulations for $\beta = 0.01, 0.5$, and for the supercritical case we compute with $\beta = 2$. The linear stability regions for these values of $\beta$, along with the critical case $\beta = 1$, are shown in figure \ref{gamma=2_stability} with the maximum allowable electric field strength $\gamma = 2$. This value of $\gamma$ gives unstable wavenumbers for all values of $\beta > 0$, hence the dynamics for small subcritical Reynolds numbers are nontrivial on sufficiently large domains. 
Figure \ref{gamma=2_stability}(a) has a relatively small value $\beta=0.01$ and shows distinct behaviour
from the other cases in panels (b)-(d); there are regions of linear stability (no unstable modes depicted with white)
in between regions of linear instability. This behaviour is not due to sub-criticality as can be seen
from the results in figure  \ref{gamma=2_stability}(b) for $\beta = 0.5$.
As many as a total of 5 modes have been computed and as expected the band of instability increases as $\nu_1$
and $\nu_2$ decreases (analogous to the domain size increasing).
Note also that in the figure, due to the symmetries of the dispersion relation \eqref{sk}, 
we count the quartet of unstable modes $(k_1,k_2)$, $(k_1, -k_2)$, $(-k_1,k_2)$, $(-k_1,-k_2)$ as one, 
with obvious special cases when either $k_1$ or $k_2$ are zero.
Regions in parameter space where solutions of (\ref{illposed2delectric2}) decay to the trivial zero solution can be obtained analytically along with bounds on the decay rates. This can be achieved using estimates on the $x$-average of solutions to (\ref{illposed2delectric2}), a Poincar\'{e}-Wirtinger inequality, and properties of the non-local operator found in Appendix \ref{PropertiesofRappendix}. However, in contrast to the one-dimensional case (see
\cite{tseluiko2006wave}), the parameter regions and the decay rate bounds obtained by these methods are not sharp, so we do not present these results here. For subcritical Reynolds number flows with the condition that the electric field strength is sufficiently weak, $\gamma < 2 (1-\beta)^{1/2}$, all Fourier modes are linearly stable for any choice of length parameters. Numerical results suggest that we have decay of all initial conditions to the zero solution for this case.

\subsection{Numerical method} 

We now move on to a numerical study of (\ref{illposed2delectric2}) on $Q$-periodic domains for which we will use the usual Fourier series representation of our solution,
\begin{equation} \eta(x,t) = \sum_{\bm{k}\in\mathbb{Z}^2} \eta_{\bm{k}}(t) e^{i \bm{\tilde{k}}\bm{\cdot} \bm{x}}. \end{equation}
We denote the norm and inner product on the space $L^2 = L_{\text{per}}^2(Q)$ as
\begin{equation} | \eta |_2= \left(\int_Q \eta^2 \; \mathrm{d}\bm{x}\right)^{1/2} = |Q|^{1/2} \left(\sum_{\bm{k}\in\mathbb{Z}^2} |\eta_{\bm{k}}|^2 \right)^{1/2},\quad \langle \eta , u \rangle_2 =  \int_Q \eta u \; \mathrm{d}\bm{x} = |Q| \sum_{\bm{k}\in\mathbb{Z}^2} \eta_{\bm{k}} u_{-\bm{k}},\end{equation} 
respectively, where $|Q| = L_1L_2$. We utilise a second-order implicit--explicit backwards differentiation formula (BDF) which belongs to a family of numerical schemes constructed by Akrivis and Crouzeix \cite{akrivis2004linearly} for a class of nonlinear parabolic equations under appropriate assumptions on the linear and nonlinear terms. They considered evolution equations of the form 
\begin{equation}\eta_t + \mathcal{A}\eta = \mathcal{B}(\eta),\end{equation}
where $\mathcal{A}$ is a positive definite, self-adjoint linear operator, and $\mathcal{B}$ is a nonlinear operator which satisfies a local Lipschitz condition. It was shown that these numerical schemes are efficient, convergent and unconditionally stable. For our consideration of (\ref{illposed2delectric2}) we have
\begin{equation}\label{mathcalABdefn} \mathcal{A}\eta = (\beta - 1) \eta_{xx} -  \eta_{yy} - \gamma \Delta \mathcal{R}(\eta) + \Delta^2 \eta  + c \eta, \qquad  \mathcal{B}(\eta) = - \eta\eta_x + c\eta,\end{equation}
where the constant $c$ is chosen to ensure that $\mathcal{A}$ is positive definite. The linear operator $\mathcal{R}$ is self-adjoint in $L^2$ (see Appendix \ref{PropertiesofRappendix}), thus $\mathcal{A}$ is also a self-adjoint linear operator. It can be shown that (see Appendix \ref{appendixnumericsposdef}) to ensure that $\mathcal{A}$ is positive definite, it is sufficient to take
\begin{equation}c > \frac{1}{2} \left[ \left(|\beta - 1|+ \gamma^2\right)^2 + \left(1 + \gamma^2\right)^2 \right].\end{equation}
The local Lipschitz condition for the nonlinear operator $\mathcal{B}$ is proved in \cite{akrivislinearly}, therefore the linearly implicit methods derived in \cite{akrivis2004linearly} are good candidates and are used for our problem.

Let $H^n$ be the approximation of the solution $\eta$ at time $n \Delta t$ for time step $\Delta t$ and $n\in \mathbb{N}$ obtained by splitting the spatial domain $Q$ into $M\times N$ equidistant points, and let $\tilde{\mathcal{A}}$ and $\tilde{\mathcal{B}}$ be the discretisations of $\mathcal{A}$ and $\mathcal{B}$ respectively. Taking $H^0$ as the discretisation of the initial condition $\eta_0$, we employ one step of the implicit Euler method as a starting approximation,
\begin{equation}\label{Eulerstep1}H^1 + \Delta t  \tilde{\mathcal{A}} H^1 = H^0 + \Delta t \tilde{\mathcal{B}}(H^0),\end{equation}
and then use the second-order implicit--explicit BDF scheme given by
\begin{equation}\label{BDFstep1}\frac{3}{2} H^{n+2} + \Delta t \tilde{\mathcal{A}} H^{n+2} = 2 H^{n+1} - \frac{1}{2} H^n + 2\Delta t \tilde{\mathcal{B}}(H^{n+1}) - \Delta t\tilde{\mathcal{B}}(H^n).\end{equation}
We take the discrete Fourier transform of (\ref{Eulerstep1}) and (\ref{BDFstep1}), denoted by $\mathcal{F}$, and solve the resulting equations in Fourier space. Let $\widehat{\mathcal{A}}$ be the discretisation of the operator $\mathcal{A}$ in Fourier space, it is a matrix operator with
\begin{equation}\widehat{\mathcal{A}}_{\bm{k}} = - (\beta - 1) \tilde{k}_1^2 + \tilde{k}_2^2 - \gamma  \left(\tilde{k}_1^2 + \tilde{k}_2^2\right)^{3/2} + \left(\tilde{k}_1^2 + \tilde{k}_2^2\right)^2 + c\end{equation}
so that
\begin{equation}\mathcal{F}(\tilde{\mathcal{A}}(H^n))_{\bm{k}} = \widehat{\mathcal{A}}_{\bm{k}} \widehat{H}^n_{\bm{k}}\end{equation}
where $\widehat{H}^n$ is the discrete Fourier transform of $H^n$. The discrete Fourier transform of the nonlinear operator $\mathcal{B}$ is given by
\begin{equation}\mathcal{F}(\tilde{\mathcal{B}}(H^n))_{\bm{k}} = - \frac{i \tilde{k}_1}{2} \mathcal{F}((H^n)^{2})_{\bm{k}} + c \widehat{H}^n_{\bm{k}}.\end{equation}
Taking the Fourier transform of the equations, we obtain the following for the implicit Euler step,
\begin{equation}\widehat{H}_{\bm{k}}^1  =\frac{ \widehat{H}_{\bm{k}}^0 + \Delta t \mathcal{F}(\tilde{\mathcal{B}}(H^0))_{\bm{k}}}{1 + \Delta t  \widehat{\mathcal{A}}_{\bm{k}} },\end{equation}
and for the second-order BDF steps,
\begin{equation} \widehat{H}_{\bm{k}}^{n+2} = \frac{4 \widehat{H}_{\bm{k}}^{n+1} -  \widehat{H}_{\bm{k}}^n + 4\Delta t  \mathcal{F}(\tilde{\mathcal{B}}(H^{n+1}))_{\bm{k}} - 2 \Delta t \mathcal{F}(\tilde{\mathcal{B}}(H^n))_{\bm{k}}}{3 + 2\Delta t \widehat{\mathcal{A}}_{\bm{k}}}.\end{equation}
The initial conditions with zero average used in our numerical simulations are
\begin{align}\nonumber \eta_0(x,y) =  \sum_{| \bm{k} |_{\infty}=1}^{20} & \left[ a_{\bm{k}} \cos\left(\tilde{k}_1 x + \tilde{k}_2 y \right) + b_{\bm{k}} \sin\left(\tilde{k}_1 x + \tilde{k}_2 y\right) \right. \\
&  \left. \quad\;\; + c_{\bm{k}} \cos\left(\tilde{k}_1 x - \tilde{k}_2 y\right) + d_{\bm{k}} \sin\left(\tilde{k}_1 x - \tilde{k}_2 y\right) \right] \end{align}
where the coefficients $a_{\bm{k}}, b_{\bm{k}}, c_{\bm{k}}$ and $d_{\bm{k}}$ are pseudorandom numbers in the range $[-0.05,0.05)$.

\subsection{Numerical results}

We do not carry out an exhaustive computational study of the dynamics as the dimensions $L_1$ and $L_2$
vary independently, due to the large number of runs required producing a significant amount of data to be analyzed.
Instead, we
restrict our attention to square periodic domains by setting $L_1 = L_2 = L$, or equivalently $\nu_1 = \nu_2 = \nu$. 
For subcritical Reynolds numbers we take $\beta = 0.01$ and $\beta = 0.5$ and as noted previously, these have very different linear stability regions as seen in figure \ref{gamma=2_stability}(a)-(b). 
For supercritical Reynolds numbers (the dynamics are non-trivial even in the absence of a field $\gamma=0$),
we take $\beta = 2$ and provide a qualitative description of the dynamics as $\gamma$ is increased. 
We will examine the attractor windows of dynamical behaviours in the length parameter $L$, in particular obtaining wave formations which are not dominated by one-dimensional behaviour. To provide a qualitative description of solutions to (\ref{illposed2delectric2}) and the nature of the attractor, we employ a number of data analysis tools. 
We rely predominantly on the $L^2$-norm, a measure of the solution energy, as a diagnostic. 
From this we construct the phase plane diagram for the energy, plotting the $L^2$-norm against its time derivative. To construct the Poincar\'{e} energy return map, we find the sequence of times $\{ t_n \}_{n = 1}^{N}$ for which the $L^2$-norm is at a minimum for a given finite time interval that can be very large. 
We then plot the points $(E_n, E_{n+1})$ where $E_n$ is the $L^2$-norm at time $t_n$. The two-dimensionality of solutions to (\ref{illposed2delectric2}) is quantified by studying the time-averaged power spectrum of solutions, given by
\begin{equation}S(\bm{k}) = |Q| \lim_{T\rightarrow\infty}\frac{1}{T}\int_0^T |\eta_{\bm{k}}|^2\; \mathrm{d}t\end{equation}
for each $\bm{k}\in\mathbb{Z}^2$. In practice we approximate $S(\bm{k})$ by 
\begin{equation}\overline{S}(\bm{k}) = \frac{|Q|}{T_2-T_1}\int_{T_1}^{T_2} |\eta_{\bm{k}}|^2\; \mathrm{d}t \end{equation}
where $0\ll T_1\ll T_2$ are two large times. Any activity in the mixed Fourier modes for solutions in the attractor will be made apparent with this diagnostic; if the time-averaged power spectrum is restricted to the $(k,0)$-modes then we will call it one-dimensional, otherwise it is called two-dimensional. The integration times used were at least $10^3$ time units, and Fourier modes of magnitude as small as $10^{-15}$ were retained. The time steps used for numerical simulations are $10^{-4}$ and smaller; for larger values of $\beta$ and $\gamma$, smaller time steps are required to obtain good convergence (see \cite{akrivislinearly} for a convergence analysis of the same scheme applied to the Kuramoto--Sivashinsky equation (\ref{KSE2dNepo})). It is worthwhile to question whether there are any issues associated with performing numerical simulations at the critical electric field strength $\gamma = 2$. From the form of the nonlinearity in (\ref{illposed2delectric2}), the problem for the transverse modes ($y$-modes) is linear and decouples completely. For $\gamma = 2$, there exist discrete values of $L_2$ at which the transverse modes are neutrally stable, otherwise these are always damped, and so the dynamical behaviour we observe at the endpoint $\gamma=2$ is not a special case, but can be found for $\gamma$ slightly less than $2$. We also note that numerical simulations were also performed for $\gamma > 2$ and as predicted by the linear theory, blow-ups are observed for some domain dimensions; this is 
not surpring given that the transverse mode problem decouples as discussed above. 

In the presentation of results that follows we use the following key for the attractor behaviour: 
\begin{enumerate}[(i)]
\item {\textrm{ }}$Z$ denotes an attractor consisting of the trivial zero solution.
\item {\textrm{ }}$D_{(k_1,k_2)}$ denotes a modal attractor (steady or travelling) in which solutions are dominated by the integer multiples of the $(k_1,\pm k_2)$-mode, for example $D_{(2,0)}$ is an attractor of bimodal states in the streamwise direction.
\item {\textrm{ }}$TP$ denotes a time-periodic attractor, more specific details will be elaborated on
where appropriate.
\item {\textrm{ }}$A$ denotes a range of attractors with complicated dynamical behaviour, including period doubling bifurcations, multimodal steady or travelling waves, time-periodic/quasiperiodic attractors, periodic bursting and chaotic attractors
\end{enumerate}

For the attractors denoted by $TP$ or $A$, the subscript $1$ or $2$ indicates whether the attractor dynamics are dominated by one-
or two-dimensional behaviour. It is important to note that due to the Galilean transformation (\ref{rescale100}) that is used to remove the advective term, all steady states correspond to travelling waves in the original frame of reference.

\subsubsection{Small subcritical Reynolds number, $\beta = 0.01$}

\begin{figure}\setcounter{figure}{2}
\caption{Schematic of the attractors for $\beta = 0.01$, $\gamma =2$. } \label{beta=0_01attractors}
\begin{tikzpicture}[scale=1.5]

\draw[-] (0, 0) -- (7,0) ;

\draw[-] (0, -0.15) node[below] {$\;L =\;\;\; 5.7 \quad\quad\quad$} -- (0,0.15);

\draw (0, -0.45) node[below] {$\;\quad\nu = \;\; 1.22 \quad\quad\quad\;\;\;$};

\draw(0.5, 0.1) node[above] {$D_{(1,0)}$};

\draw[-] (1, -0.15) node[below] { $ 7.0 $} -- (1,0.15);

\draw (1,  -0.45) node[below] {$  0.81 $} ;

\draw(1.5, 0.1) node[above] {$ Z $};

\draw[-] (2, -0.15) node[below] {$ 8.3 $} -- (2,0.15);

\draw (2,  -0.45) node[below] {$ 0.57 $} ;

\draw(2.5, 0.1) node[above] {$ D_{(1,1)}  $};

\draw[-] (3, -0.15) node[below] {$  9.6 $} -- (3,0.15);

\draw (3,  -0.45) node[below] {$ 0.43  $} ;

\draw(3.5, 0.1) node[above] {$ Z $};

\draw[-] (4, -0.15) node[below] {$ 11.4 $} -- (4,0.15);

\draw (4,  -0.45) node[below] {$ 0.30  $} ;

\draw(4.5, 0.1) node[above] {$ A_2 $};

\draw[-] (5, -0.15) node[below] {$ 13.6 $} -- (5,0.15);

\draw (5,  -0.45) node[below] {$ 0.21  $} ;

\draw(5.5, 0.1) node[above] {$ D_{(1,2)} $};

\draw[-] (6, -0.15) node[below] {$ 14.7 $} -- (6,0.15);

\draw (6,  -0.45) node[below] {$ 0.18  $} ;

\draw(6.5, 0.1) node[above] {$  A_2 $};

\draw[-] (7, -0.15) -- (7,0.15);

\draw[-] (0, -1.5) -- (3,-1.5) ;

\draw[-] (0, -1.65) node[below] {$\;L =\;\; 15.4 \quad\quad\quad$} -- (0,-1.35);

\draw[-] (0, -1.95) node[below] {$\;\quad\nu = \;\; 0.17 \quad\quad\quad\;\;\;$};

\draw(0.5, -1.4) node[above] {$ Z $};

\draw[-] (1, -1.65) node[below] {$ 16.6 $} -- (1,-1.35);

\draw (1, -1.95) node[below] {$ 0.14 $} ;

\draw(1.5,-1.4) node[above] {$D_{(2,2)} $};

\draw[-] (2, -1.65) node[below] {$ 18.7 $} -- (2,-1.35);

\draw (2, -1.95) node[below] {$ 0.11   $} ;

\draw(2.5, -1.4) node[above] {$ A_2 $};

\end{tikzpicture}
\end{figure}
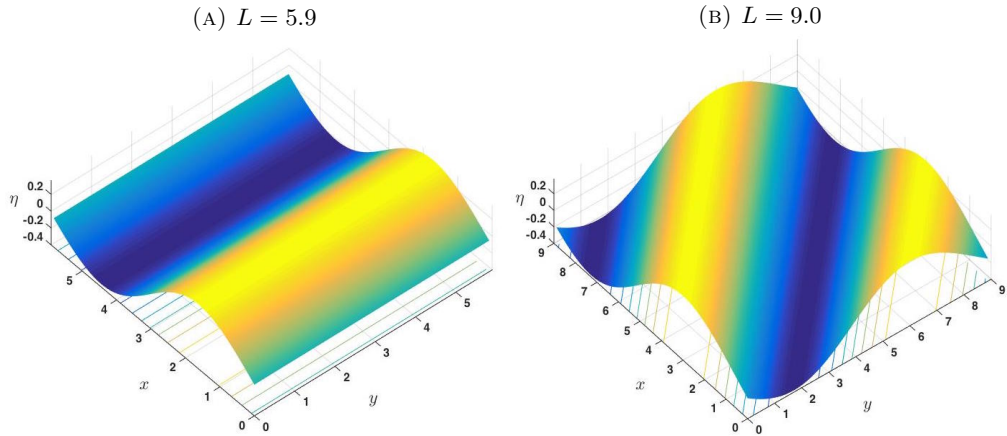

For $\beta = 0.01$ and $\gamma < 2\sqrt{0.99} \approx 1.9900$, the linear theory and numerics predict that we have decay of all solutions to zero for arbitrary initial conditions, and so we concentrate on the case of $\gamma = 2$. The linear stability regions for this choice of $\gamma$ are depicted in figure \ref{gamma=2_stability}(a), and it can be seen that along the line along which the numerical results are obtained ($\nu_1 = \nu_2$) there is initially alternation between linear stability and instability. Figure \ref{beta=0_01attractors} was constructed from a large number of numerical experiments to collect a broad qualitative description of the solution attractors. As $L$ increases, the $(1,0)$-mode becomes linearly unstable first at $L = 5.7$, and an attractor of one-dimensional unimodal steady states and travelling waves as is observed (these are analogous to other Kuramoto--Sivashinsky-type equations). 
An example of such a profile from this unimodal $D_{(1,0)}$ window is given in figure \ref{D1andD11profiles}(a). 
Increasing $L$ further to $7.0$, the $(1,0)$-mode then becomes stable again and all initial conditions are attracted to the zero solution -
see the schematic in figure \ref{beta=0_01attractors}. This process is repeated when the $(1,1)$-mode is destabilised at $L=8.3$, and diagonal unimodal steady states and travelling waves are observed, dominated by the $(k,k)$-modes or $(k,-k)$-modes depending on the initial condition. Figure \ref{D1andD11profiles}(b) shows an example of a solution profile of type $D_{(1,1)}$ in this attractor. 
As previously, a region of linear stability in all Fourier modes is then reached at $L=9.6$ and this persists until $L=11.4$, approximately. 
Increasing $L$ further we find an increasingly complicated sequence of attractors. 
For $L$ between $11.4$ and $15.4$, at most three modes are linearly unstable, the $(2,0)$, $(1,2)$ and $(2,1)$-modes. 
Initially, increasing $L$ above $11.4$ we see a time-periodic and quasi-time-periodic attractors with homoclinic bursting behaviour, 
where the profile switches between an odd pair (under the parity transformation) of bimodal states through a short two-dimensional pulse transition period (see supplementary Movie 1 available at \url{https://youtu.be/yZKc7qbwPKM} for a time-periodic solution). 
Beyond $L=13.6$, the $(1,2)$-mode dominates and we observe a window of the attractor $D_{(1,2)}$. 
For $L$ above $14.7$, we mostly observe attractors with homoclinic bursting behaviours with long burst times. All modes become linearly stable again at $L=15.4$, 
and non-trivial behaviour is not found until
$L=16.6$ when the $(2,2)$-mode becomes unstable and a $D_{(2,2)}$ solution emerges initially. 
For $L$ above $18.7$, the dynamics become increasingly complicated (see supplementary Movie 2 available at \url{https://youtu.be/KgQb6xGctcU} for a quasi-time-periodic solution exhibiting homoclinic bursting behaviour for $L = 18.85$, 
where the interface undergoes transitions between a pulse state and a ``snaking" transverse wave). Finally,
fully chaotic behaviour is found for sufficiently large $L$.

\begin{figure}
\caption{Profiles of solutions in $D_{(1,0)}$ and $D_{(1,1)}$ for $\beta = 0.01$, $\gamma = 2$.} \label{D1andD11profiles}
\begin{subfigure}{2.6in}
\caption{$L = 5.9$} 
\includegraphics[width=2.6in]{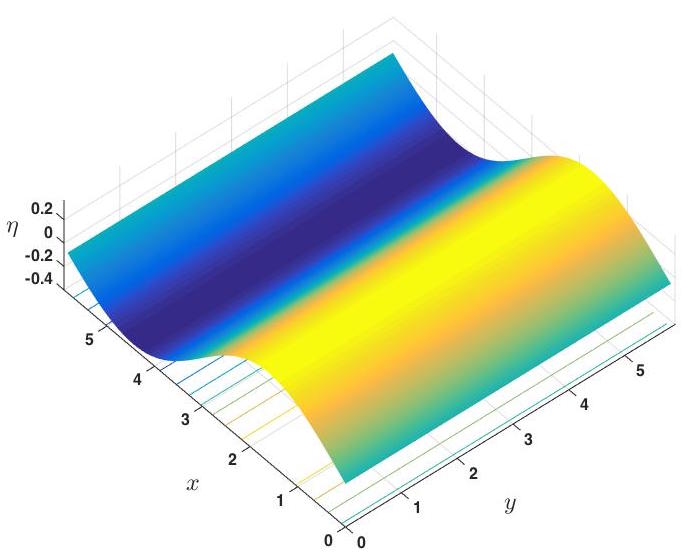}
\end{subfigure}
\begin{subfigure}{2.6in}
\caption{$L = 9.0$} 
\includegraphics[width=2.6in]{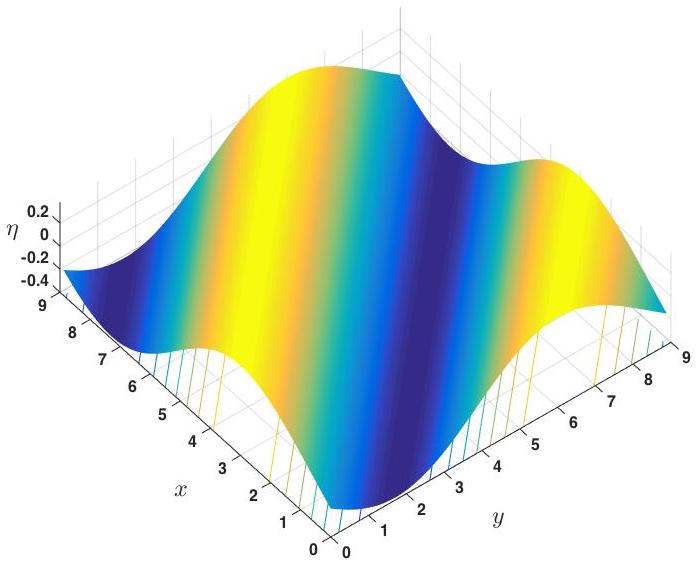}
\end{subfigure}
\end{figure}

\subsubsection{Moderate subcritical Reynolds number, $\beta = 0.5$}

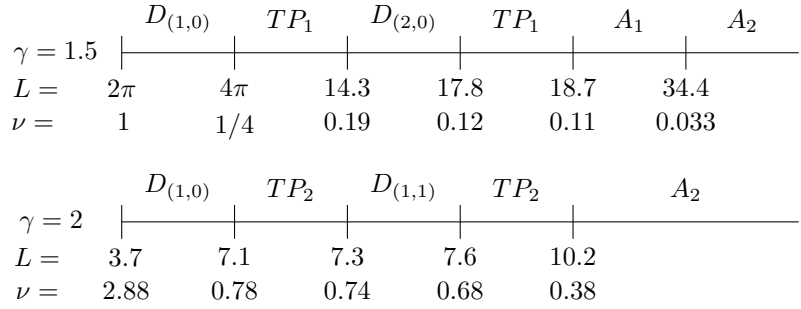
\begin{figure}\setcounter{figure}{4}
\caption{Schematic of the attractors (not drawn to scale) for $\beta = 0.5$, $\gamma = 1.5,\; 2$.} \label{beta=0_5attractors}

\begin{tikzpicture}[scale=1.5]

\draw[-] (0, 0) node[left] {$\gamma = 1.5\;\;$} -- (6,0) ;

\draw[-] (0, -0.15) node[below] {$\;L =\quad\;\; 2\pi \quad\quad\quad\;\;\;$} -- (0,0.15);

\draw (0, -0.45) node[below] {$\nu =\quad\;\;\;\; 1 \quad\quad\quad\;\;\;$};

\draw(0.5, 0.1) node[above] {$D_{(1,0)}$};

\draw[-] (1, -0.15) node[below] {$ 4\pi $} -- (1,0.15);

\draw (1, -0.45) node[below] {$ 1/4 $} ;

\draw(1.5, 0.1) node[above] {$TP_1$};

\draw[-] (2, -0.15) node[below] {$14.3 $} -- (2,0.15);

\draw (2, -0.45) node[below] {$ 0.19 $} ;

\draw(2.5, 0.1) node[above] {$ D_{(2,0)} $};

\draw[-] (3, -0.15) node[below] {$ 17.8 $} -- (3,0.15);

\draw (3, -0.45) node[below] {$ 0.12 $} ;

\draw(3.5, 0.1) node[above] {$ TP_1 $};

\draw[-] (4, -0.15) node[below] {$ 18.7 $} -- (4,0.15);

\draw (4, -0.45) node[below] {$ 0.11 $} ;

\draw(4.5, 0.1) node[above] {$ A_1  $};

\draw[-] (5, -0.15) node[below] {$ 34.4 $} -- (5,0.15);

\draw[-] (5, -0.45) node[below] {$ 0.033 $};

\draw(5.5, 0.1) node[above] {$ A_2  $};

\draw[-] (0, -1.5) node[left] {$\gamma = 2\;\;\;\;$} -- (6,-1.5) ;

\draw[-] (0, -1.65) node[below] {$L =\;\;\quad  3.7 \qquad\;\quad$} -- (0,-1.35);

\draw[-] (0, -1.95) node[below] {$\nu =\;\;\;\;\; 2.88 \quad\;\;\;\;\;\;\;$};

\draw(0.5, -1.4) node[above] {$D_{(1,0)}$};

\draw[-] (1,-1.65) node[below] {$ 7.1$} -- (1,-1.35);

\draw (1, -1.95) node[below] {$ 0.78$} ;

\draw(1.5,-1.4) node[above] {$ TP_2 $};

\draw[-] (2, -1.65) node[below] {$ 7.3 $} -- (2,-1.35);

\draw (2, -1.95) node[below] {$ 0.74  $} ;

\draw(2.5, -1.4) node[above] {$D_{(1,1)}  $};

\draw[-] (3, -1.65) node[below] {$7.6 $} -- (3,-1.35);

\draw (3, -1.95) node[below] {$ 0.68 $} ;

\draw(3.5, -1.4) node[above] {$ TP_2 $};

\draw[-] (4, -1.65) node[below] {$10.2   $} -- (4,-1.35);

\draw (4, -1.95) node[below] {$ 0.38  $} ;

\draw(5, -1.4) node[above] {$ A_2 $};

\end{tikzpicture}

\end{figure}

Having considered small inertia effects, we now turn to larger values of $\beta$ but still in the subcritical regime.
We pick $\beta=0.5$, in which case
linear theory and numerical solutions predict decay of all initial conditions to the trivial 
zero solution for $\gamma < \sqrt{2} \approx 1.4142$. Thus we will investigate the cases $\gamma = 1.5$ and $2.0$ -
the linear stability regions for $\beta=0.5$, $\gamma = 2$ are displayed in figure \ref{gamma=2_stability}(b). 
The figure shows clearly that in contrast to the smaller inertia case $\beta=0.01$, there are no regions of stability after the first mode becomes linearly unstable,
and hence non-trivial dynamics are expected throughout as $L$ increases.
This is confirmed by the results of figure \ref{beta=0_5attractors} which depicts the most attracting states as $L$ increases for $\gamma=1.5$ and $2.0$.

For $\gamma = 1.5$, the zero solution loses stability to the $(1,0)$-mode when $L$ exceeds 
$2\pi$, and a window of unimodal states $D_{(1,0)}$ emerges. 
Note that according to linear theory the $(1,0)$-mode becomes stable at $L = 4\pi$ and for $L>4\pi$ the $(2,0)$-mode loses stability. At $L=4\pi$ we find a Hopf bifurcation with a time-periodic spatially one-dimensional $TP_1$ solution emerging
until $L=14.3$ - these solutions are homoclinic bursts with the long-lived $D_{(2,0)}$ solutions undergoing time periodic oscillations through unimodal $D_{(1,0)}$ states. The next attractor window, $14.3<L<17.8$, contains bimodal $D_{(2,0)}$ states that in turn lose stability via a Hopf bifurcation
to time periodic solutions (no homoclinic bursting) in the window $17.8<L<18.7$. The strong one-dimensionality
persists in the window $18.7<L<34.4$ and complex dynamics including
trimodal steady states and chaotic bursting are found.
Beyond this, the mixed modes remain active in chaotic solutions, and are characterised by the presence of small deformations on the usual cellular one-dimensional chaotic profiles.

The dynamics for $\gamma = 2$ are much more interesting. 
As mentioned above, as the strength of the destabilising electric field is increased, the more complicated dynamics 
appear for lower values of $L$. There is also a change in the attractor windows observed, with 
increased and persistent two-dimensionality due to the electric field intensifying 
the instability in the mixed modes. As summarised in figure \ref{beta=0_5attractors}, beyond $L = 3.7$ 
we observe a window of unimodal states as before, 
but the next window between $L=7.1$ and $L=7.3$ exhibits two-dimensional time-periodic behaviour 
(see supplementary Movie 3 available at \url{https://youtu.be/PQa9IOvwiZU}). 
The time periodic solutions become less attractive as $L$ increases, and in the window $7.3<L<7.6$ they give way to diagonal modal $D_{(1,1)}$ states similar to those obtained 
for $\beta = 0.01$, $\gamma = 2$ shown in figure \ref{D1andD11profiles} (b). Between $L=7.7$ and $10.2$, we observe a window of two-dimensional time-periodic homoclinic bursting behaviour (labeled $TP_2$ on figure \ref{beta=0_5attractors}), while for
$L=10.2$ onwards we find a range of very interesting fully two-dimensional solutions before the onset of chaos. Several solutions from this range are depicted in figure \ref{b=0_5_g=2}. Panel (a) shows the profile of a 
quasi-periodic in time solution at $L=19.0$; the underlying pulse structures travel in the $x$-direction and modulate weakly, 
but otherwise retain their shape and coherent details
(see supplemental Movie 4 available at \url{https://youtu.be/fdANKuioM9Q} of which figure \ref{b=0_5_g=2} (a) is a snapshot). 
Figure \ref{b=0_5_g=2}(b)-(d) 
show profiles of steady solutions at $L=19.5,\,21.0$ and $22.2$. All three of these are stable in the sense that
they are computed from initial value problems that reach steady states.
Panel (b) corresponds 
to a solution in the attractor $D_{(2,3)}$, while panel (c) displays a rather unusual ``snaking" steady state (reminiscent of the 
quiescent state of the homoclinic bursting shown in the supplemental Movie 2 available at \url{https://youtu.be/KgQb6xGctcU}).
The profile in panel (d) is found to be similar to that of panel (b) but has a pulse
disturbing the structure; the pulse has dimensions analogous to those in panel (a) and hence we can conclude
that there is an interplay between different attractors producing quite intricate two-dimensional interfacial steady states.

\begin{figure}
\caption{Window $A_2$, $\beta = 0.5$, $\gamma = 2$} \label{b=0_5_g=2}
\begin{subfigure}{2.6in}
\caption{ $ L=19.0$} 
\includegraphics[width=2.6in]{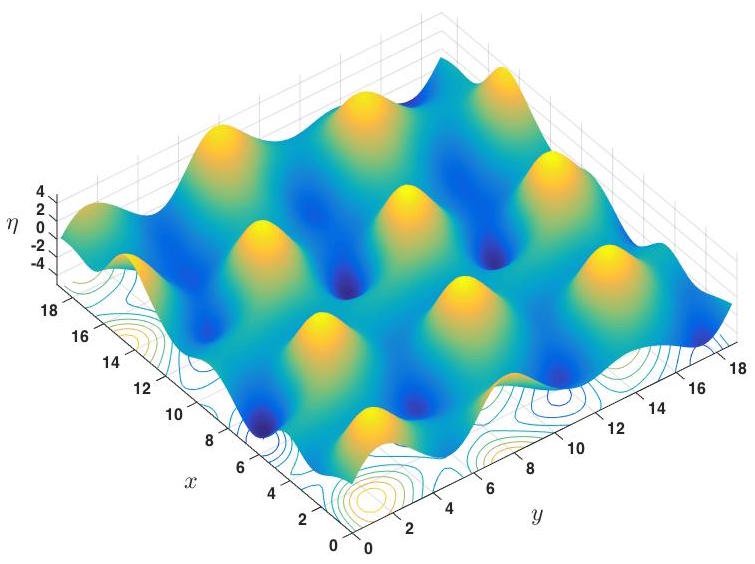}
\end{subfigure}
\begin{subfigure}{2.6in}
\caption{ $ L=19.5$} 
\includegraphics[width=2.6in]{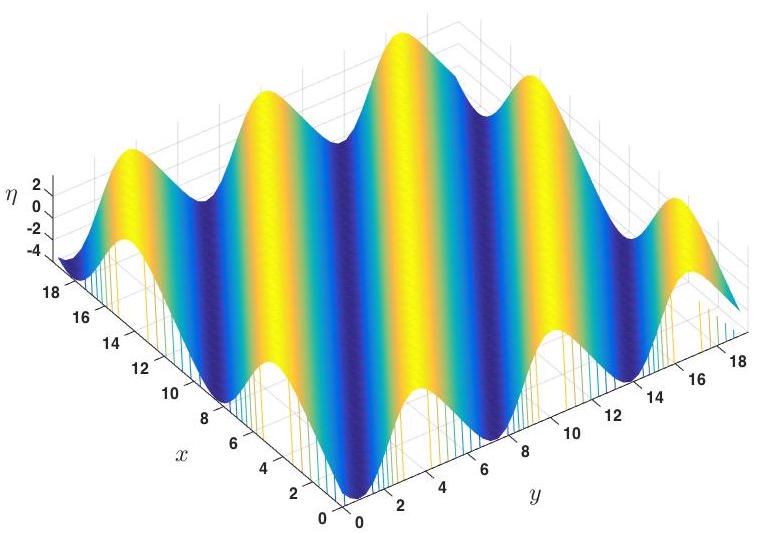}
\end{subfigure}
\begin{subfigure}{2.6in}
\caption{ $ L=21.0$} 
\includegraphics[width=2.6in]{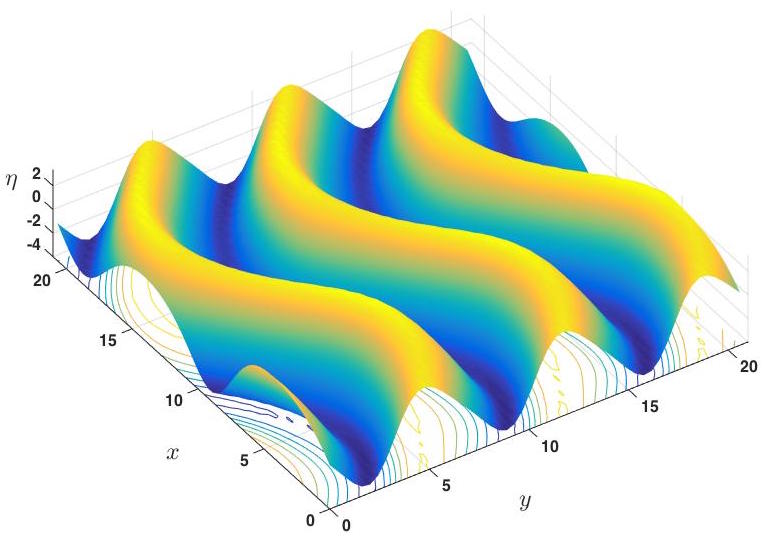}
\end{subfigure}
\begin{subfigure}{2.6in}
\caption{ $ L=22.2$} 
\includegraphics[width=2.6in]{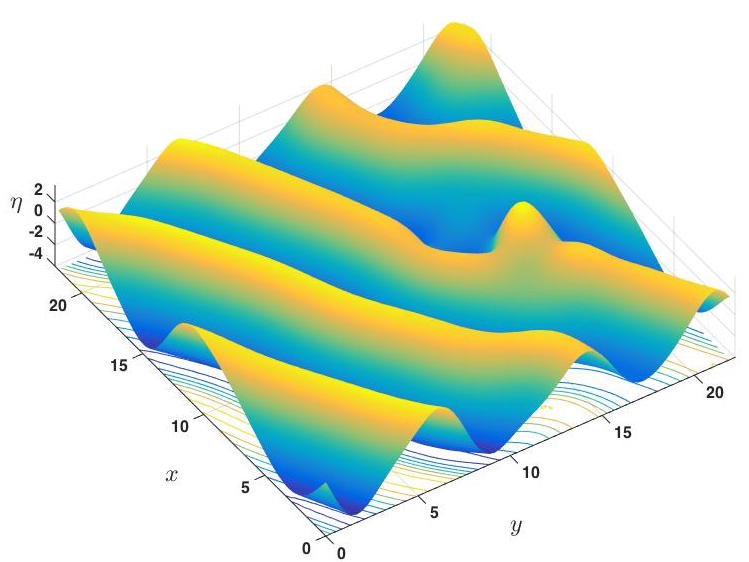}
\end{subfigure}
\end{figure}

\subsubsection{Supercritical Reynolds number, $\beta = 2$}

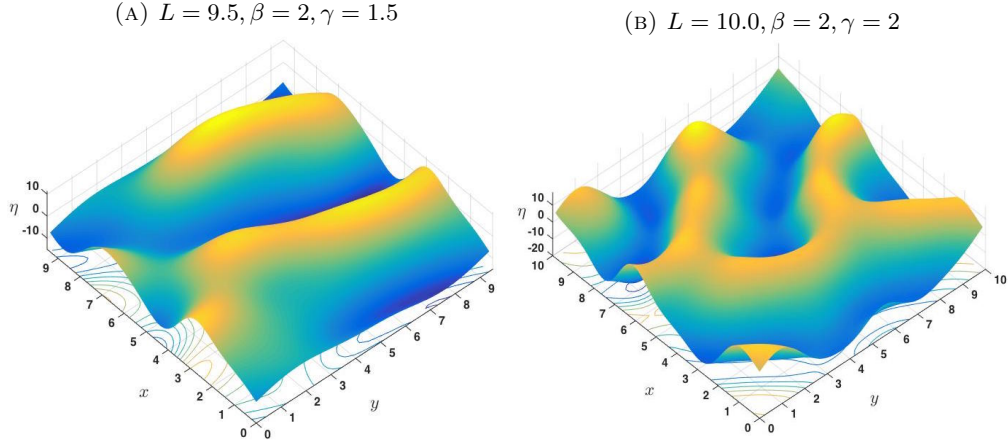
\begin{figure}\setcounter{figure}{6}
\caption{Schematic of the attractors for $\beta = 2$, $\gamma = 0,\; 0.5,\; 1, \; 1.5, \; 2$.} \label{beta=2attractors}

\begin{tikzpicture}[scale=1.5]

\draw[-] (0, 0) node[left] {$\gamma = 0\;\;\;\;\;$} -- (5,0) ;

\draw[-] (0, -0.15) node[below] {$\;L =\quad\;\; 6.3 \quad\quad\quad\;\;\;$} -- (0,0.15);

\draw (0, -0.45) node[below] {$\nu =\quad\;\;\;\; 1  \quad\quad\quad\;\;\;$};

\draw(0.5, 0.1) node[above] {$ D_{(1,0)} $};

\draw[-] (1, -0.15) node[below] {$ 13.1  $} -- (1,0.15);

\draw (1, -0.45) node[below] {$ 0.23 $} ;

\draw(1.5, 0.1) node[above] {$ TP_1 $};

\draw[-] (2, -0.15) node[below] {$  14.9 $} -- (2,0.15);

\draw (2, -0.45) node[below] {$ 0.18  $} ;

\draw(2.5, 0.1) node[above] {$ D_{(2,0)} $};

\draw[-] (3, -0.15) node[below] {$ 17.3  $} -- (3,0.15);

\draw (3, -0.45) node[below] {$ 0.13 $} ;

\draw(3.5, 0.1) node[above] {$ TP_1  $};

\draw[-] (4, -0.15) node[below] {  $17.9$ } -- (4,0.15);

\draw (4, -0.45) node[below] {$ 0.12  $} ;

\draw(4.5, 0.1) node[above] {$ A_2  $};


\draw[-] (0, -1.5) node[left] {$\gamma = 0.5\;\;$} -- (5,-1.5) ;

\draw[-] (0, -1.65) node[below] {$L =\;\;\quad 4.9  \qquad\;\;\;\;\;\;$} -- (0,-1.35);

\draw[-] (0, -1.95) node[below] {$\nu =\;\;\;\;\;\; 1.64  \quad\;\;\;\;\;\;\;\;$};

\draw(0.5, -1.4) node[above] {$ D_{(1,0)} $};

\draw[-] (1, -1.65) node[below] {$ 10.1  $} -- (1,-1.35);

\draw (1, -1.95) node[below] {$ 0.39  $} ;

\draw(1.5,-1.4) node[above] {$ TP_{1/2} $};

\draw[-] (2, -1.65) node[below] {$ 12.0 $} -- (2,-1.35);

\draw (2, -1.95) node[below] {$  0.27   $} ;

\draw(2.5, -1.4) node[above] {$ D_{(2,0)}   $};

\draw[-] (3, -1.65) node[below] {$ 13.5   $} -- (3,-1.35);

\draw (3, -1.95) node[below] {$ 0.22 $} ;

\draw(3.5, -1.4) node[above] {$ TP_1  $};

\draw[-] (4, -1.65) node[below] {$ 13.8  $} -- (4,-1.35);

\draw (4, -1.95) node[below] {$ 0.21 $} ;

\draw(4.5, -1.4) node[above] {$ A_2  $};


\draw[-] (0, -3) node[left] {$\gamma = 1\;\;\;\;\;$} -- (5,-3) ;

\draw[-] (0, -3.15) node[below] {$L =\;\;\;\;\;\; 3.9  \qquad\;\;\;\;\;\;$} -- (0,-2.85);

\draw[-] (0, -3.45) node[below] {$\nu =\;\;\;\;\;\; 2.60 \quad\;\;\;\;\;\;\;\;$};

\draw(0.5, -2.9) node[above] {$ D_{(1,0)} $};

\draw[-] (1, -3.15) node[below] {$ 8.0  $} -- (1,-2.85);

\draw (1, -3.45) node[below] {$ 0.62  $} ;

\draw(1.5,-2.9) node[above] {$ TP_{1/2} $};

\draw[-] (2, -3.15) node[below] {$ 9.6 $} -- (2,-2.85);

\draw (2, -3.45) node[below] {$  0.43   $} ;

\draw(2.5, -2.9) node[above] {$ D_{(2,0)}   $};

\draw[-] (3, -3.15) node[below] {$ 10.7   $} -- (3,-2.85);

\draw (3, -3.45) node[below] {$ 0.34 $} ;

\draw(3.5, -2.9) node[above] {$ TP_1  $};

\draw[-] (4, -3.15) node[below] {$ 10.9  $} -- (4,-2.85);

\draw (4, -3.45) node[below] {$ 0.33 $} ;

\draw(4.5, -2.9) node[above] {$ A_2  $};


\draw[-] (0, -4.5) node[left] {$\gamma = 1.5\;\;$} -- (5,-4.5) ;

\draw[-] (0, -4.65) node[below] {$L =\;\;\;\;\; 3.1  \qquad\;\;\;\;\;$} -- (0,-4.35);

\draw[-] (0, -4.95) node[below] {$\nu =\;\;\;\;\;4.11 \quad\;\;\;\;\;\;\;$};

\draw(0.5, -4.4) node[above] {$ D_{(1,0)} $};

\draw[-] (1, -4.65) node[below] {$ 6.5  $} -- (1,-4.35);

\draw (1, -4.95) node[below] {$ 0.93  $} ;

\draw(1.5,-4.4) node[above] {$ TP_{1/2} $};

\draw[-] (2, -4.65) node[below] {$ 7.9 $} -- (2,-4.35);

\draw (2, -4.95) node[below] {$  0.63   $} ;

\draw(2.5, -4.4) node[above] {$ D_{(2,0)}   $};

\draw[-] (3, -4.65) node[below] {$ 8.6   $} -- (3,-4.35);

\draw (3, -4.95) node[below] {$ 0.53 $} ;

\draw(3.5, -4.4) node[above] {$ TP_1  $};

\draw[-] (4, -4.65) node[below] {$ 8.8   $} -- (4,-4.35);

\draw (4, -4.95) node[below] {$ 0.51 $} ;

\draw(4.5, -4.4) node[above] {$ A_2  $};

  
  \draw[-] (0, -6) node[left] {$\gamma = 2\;\;\;\;\;$} -- (5,-6) ;

\draw[-] (0, -6.15) node[below] {$L =\;\;\;\;\; 2.6  \qquad\;\;\;\;\;$} -- (0,-5.85);

\draw[-] (0, -6.45) node[below] {$\nu =\;\;\;\;\;\; 5.84 \quad\;\;\;\;\;\;\;$};

\draw(0.5, -5.9) node[above] {$ D_{(1,0)} $};

\draw[-] (1, -6.15) node[below] {$ 5.0 $} -- (1,-5.85);

\draw (1, -6.45) node[below] {$ 1.58  $} ;

\draw(1.5,-5.9) node[above] {$ TP_2 $};

\draw[-] (2, -6.15) node[below] {$ 5.2  $} -- (2,-5.85);

\draw (2, -6.45) node[below] {$  1.46   $} ;

\draw(2.5, -5.9) node[above] {$  D_{(1,0)}   $};

\draw[-] (3, -6.15) node[below] {$ 5.3   $} -- (3,-5.85);

\draw (3, -6.45) node[below] {$ 1.41 $} ;

\draw(4, -5.9) node[above] {$ A_2  $};

\end{tikzpicture}
\end{figure}

For $\beta = 2$, we have non-trivial dynamics for all values of $\gamma$ and for sufficiently large domain lengths; thus, to obtain
a picture of the dynamics as the electric field increases we consider the cases $\gamma = 0,\, 0.5,\,1,\,1.5$ and $2.0$. 
The linear stability regions for the critical field strength $\gamma = 2$ (and $\beta=2$) have been given earlier
in figure \ref{gamma=2_stability}(d), which indicates that there are no islands of linear stability in $\nu_1-\nu_2$ space.
Extensive computations were undertaken to construct a solution phase diagram as before, and this is given in
Figure \ref{beta=2attractors}.
For brevity we will not go into the details of these windows, but note that on the whole the same sequence of attractors that was found for
smaller inertia is observed also for $\beta=2$ as $L$ increases, i.e.
\begin{equation}
D_{(1,0)} \rightarrow TP_{1/2} \rightarrow D_{(2,0)} \rightarrow TP_1 \rightarrow A_2,
\end{equation}
with the exception of $\gamma = 0$ and $2$. The first time-periodic window exhibits homoclinic bursting behaviour, and the dynamics transition from one- to two-dimensional within the window. The second time periodic window exhibits one-dimensional dynamics, and the time-periodicity is
not of bursting type. Note also that this sequence and pattern of windows is similar to that found in other cases 
(see figure \ref{beta=0_5attractors}, for instance, for $\beta=0.5$). For $\gamma = 0$, we do not observe a transition from one- to two-dimensional in the first time-periodic window, and for $\gamma = 2$, we observe a second window of unimodal states after the first two-dimensional time periodic window. 
All of the windows labelled $A_2$ contain the usual complicated range of dynamics, eventually entering chaotic regimes as $L$
increases further. Figure \ref{A2b=2travelandsteady} gives examples of the fully two-dimensional 
interfacial dynamics supported in the windows $A_2$; panel (a) shows the profile of a wave travelling in an oblique angle for $\gamma = 1.5$, and panel (b) shows a steady state for $\gamma = 2$.

\begin{figure}
\caption{Representative profiles from windows $A_2$ for $\beta=2.0$: (a) a travelling wave, and (b) a steady state in the window $A_2$ for the values of $\gamma=1.5$
and $2.0$ respectively. The values of $L$ are $9.5$ and $10.0$ respectively.} \label{A2b=2travelandsteady}
\begin{subfigure}{2.6in}
\caption{ $ L= 9.5,\beta = 2, \gamma = 1.5$} 
\includegraphics[width=2.6in]{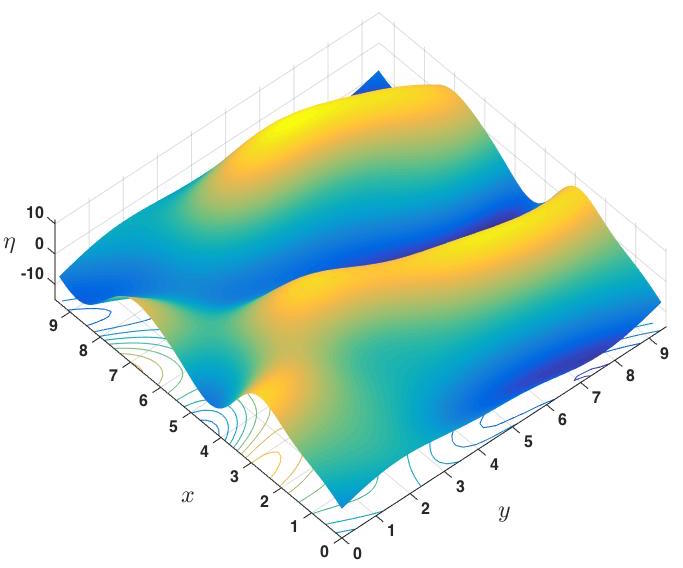}
\end{subfigure}
\begin{subfigure}{2.6in}
\caption{ $L = 10.0, \beta = 2, \gamma = 2$} 
\includegraphics[width=2.6in]{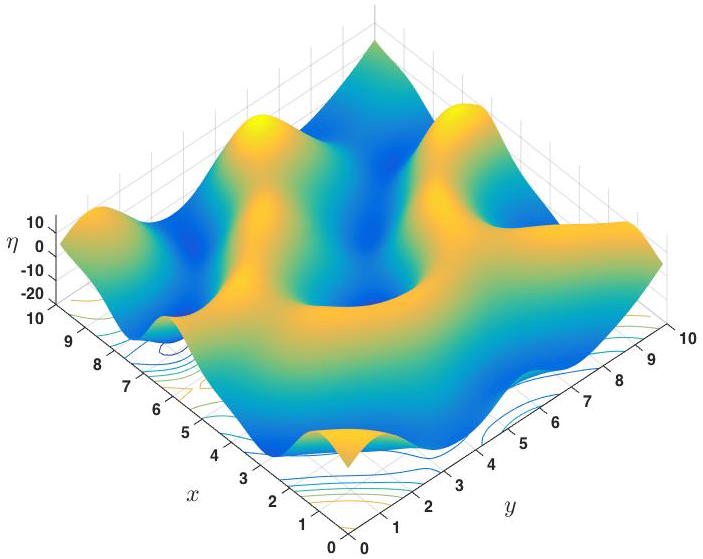}
\end{subfigure}
\end{figure}

Finally, we briefly discuss the qualitative effect of introducing an electric field to a dynamical regime that is
already chaotic. For chaotic dynamics to arise in the absence of an electric field we require a supercritical Reynolds number that already provides complex dynamics on periodic domains of sufficiently large lengths. 
We select the dimensions of the system to be $L=30$ so that chaotic dynamics are seen in the absence of a field, i.e. $\gamma=0$; figure
\ref{b=2_L=30}(a) shows a snapshot of the chaotic solution for this case and the main thing to note is that the interfacial profile remains strongly
one-dimensional throughout the evolution. In the results depicted in figures \ref{b=2_L=30}(b)-(e), the electric field parameter 
is increased to $\gamma=0.5$, $1.0$, $1.5$ and $2$ respectively. The flow remains chaotic as expected, and the snapshots shown
indicate that the field has a crucial effect in introducing two-dimensionality into the interfacial fluctuations, and also
increases the number of cellular structures, their amplitude, and hence the energy of the solutions. 
A more complete presentation of the time evolution and dynamics of solutions in this regime can be found in the supplemental Movie $5$ (available at \url{https://youtu.be/32UObKLRieM}). The movie is constructed by increasing $\gamma$
after intervals of $20$ time units, explicitly we take
\begin{equation}\gamma(t) =  \begin{cases}
0 \quad \text{if } 0 \leq t < 20, \\
1 \quad \text{if } 20 \leq t < 40,\\
2 \quad \text{if } 40 \leq t < 60.
\end{cases}
\end{equation}
We find that an increase in $\gamma$ increases the frequency of the chaotic oscillations as well as the
amplitude of the solution (the average energy increases from approximately $40$ to $100$ and then to
approximately $240$, as $\gamma$ increases from $0$ to $1$ and finally $2$ as described above). For example, in the interval $20\leq t<40$
we observe approximately seven oscillations, whereas increasing to $\gamma=2$ in the interval $40\leq t<60$ produces
roughly $20$ oscillations. These results show that even
for supercritical Reynolds numbers where there is already instability in the $x$-direction without an electric field effect, 
the transverse dynamics are non-trivial and not dominated by one-dimensional behaviour.

\begin{figure}\setcounter{figure}{8}
\caption{Profiles of solutions in the chaotic regime for $\beta = 2$, $L = 30.0$.} \label{b=2_L=30}
\begin{subfigure}{2.6in}
\caption{ $ \gamma = 0$} 
\includegraphics[width=2.6in]{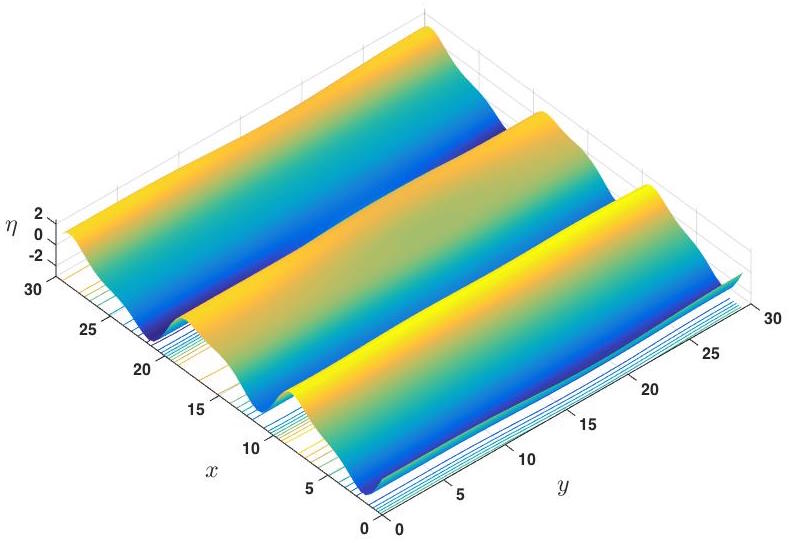}
\end{subfigure}
\begin{subfigure}{2.6in}
\caption{ $ \gamma = 0.5$} 
\includegraphics[width=2.6in]{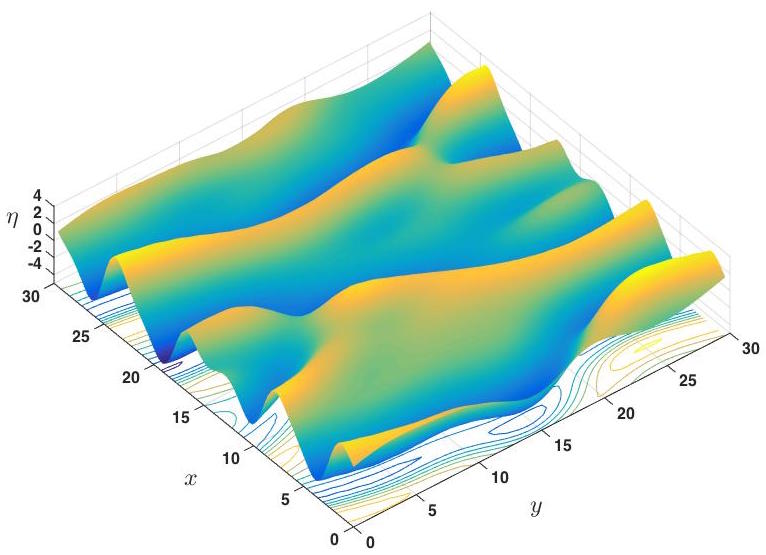}
\end{subfigure}
\begin{subfigure}{2.6in}
\caption{ $ \gamma = 1$} 
\includegraphics[width=2.6in]{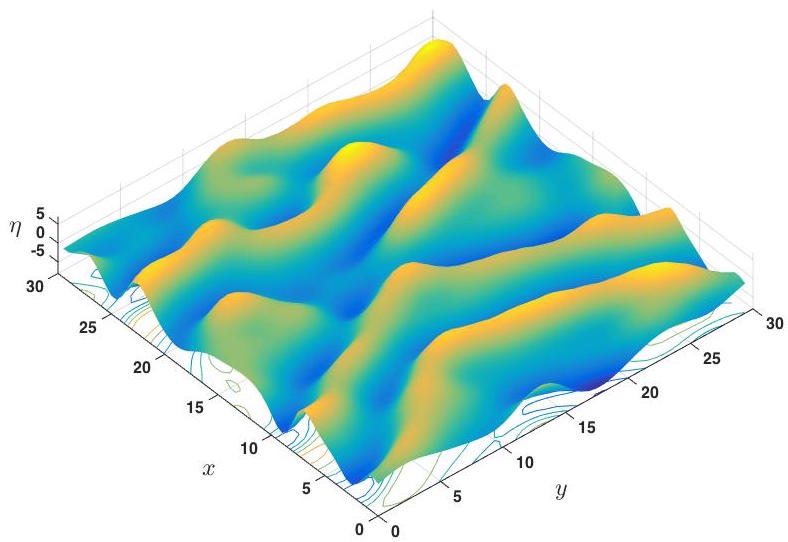}
\end{subfigure}
\begin{subfigure}{2.6in}
\caption{ $ \gamma = 1.5$} 
\includegraphics[width=2.6in]{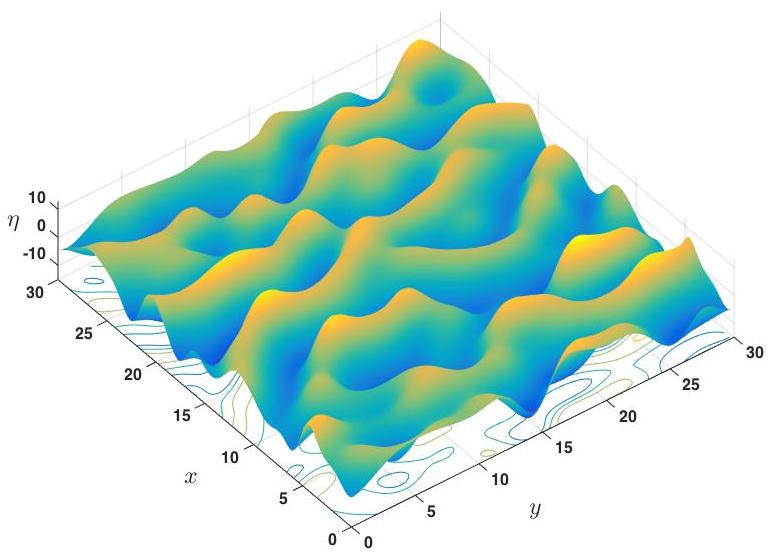}
\end{subfigure}
\begin{subfigure}{2.6in}
\caption{ $ \gamma = 2$} 
\includegraphics[width=2.6in]{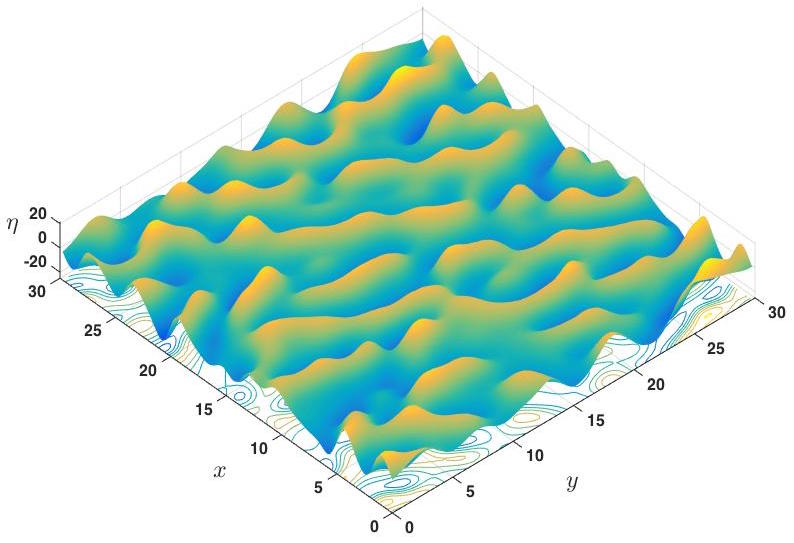}
\end{subfigure}
\end{figure}

\section{Conclusions and future directions\label{ConclusionsandFutureDirectionsSec}}

We derived a long-wave Benney model (\ref{benneyqequation1}) that describes three-dimensional long wave dynamics
of gravity driven thin film flows under the action of a normal electric field. The weakly nonlinear evolution is found to be 
governed by a Kuramoto--Sivashinsky equation with a linear non-local term (\ref{illposed2delectric2}) corresponding
to the electric field. Interestingly, solutions to this equation are bounded only if the electric field strength is below a threshold
value and the film is overlying - otherwise, unbounded exponential growth of the transverse modes cannot be prevented. The critical electric field strength
is set by the condition that all purely transverse modes are linearly stable; mixed modes can still be unstable however, and hence
produce non-trivial nonlinear three-dimensional phenomena. The present study has documented numerically a host of 
rich dynamical phenomena produced by (\ref{illposed2delectric2}) on periodic domains as the system size changes
(a more in depth study of the solution space is warranted but is beyond the scope of the present work).
We have also tried to include higher order terms of order $\delta$ (that cannon be scaled out) in the weakly nonlinear evolution in an effort
to investigate whether structural stability can be attained with bounded solutions emerging. We find that this does not happen
and in fact even more instabilities can enter resulting in enhanced ill-posedness, something that is not
unusual in gradient expansions.

An important question to pose is, what happens when the electric field strength is above critical, i.e. 
$\overline{{\Weber}} > \left( 2 \cot\theta /\overline{{\Capil}} \right)^{1/2}$.
In this case the weakly nonlinear analysis breaks down and hence we need to revert to the fully
nonlinear Benney equation (\ref{benneyqequation1}). Current work on this problem by the authors suggests
that transverse structures form that are connected by thin film regions with de-wetting being possible, for both hanging and overlying films - the results
will be presented elsewhere. In all such configurations, extensions of the weighted residual method can be undertaken
for the three-dimensional problems at hand, in order to derive equations that can be applied beyond critical in
the presence of inertia. These directions are left for future work.

Finally, it is important to point out that equation (\ref{illposed2delectric2}) supports
pattern formation phenomena and derives directly from an asymptotic analysis of the Navier--Stokes equations coupled with 
electrostatics. It is therefore of intrinsic interest as a pattern-forming two-dimensional evolution equation in analogous ways
to the Swift--Hohenberg equation \cite{Swift-H}.

\appendix

\section{Properties of the non-local operator $\mathcal{R}$\label{PropertiesofRappendix}} 

For $m\in \mathbb{R}$, let $H^m=H^m_{\text{per}}(Q)$ denote the Sobolev space of real-valued $Q$-periodic functions such that
\begin{equation} \| \eta \|_m^2 = |Q| \sum_{\bm{k}\in\mathbb{Z}^2} \left(1 + |\bm{\tilde{k}}| \right)^{2m} |\eta_{\bm{k}}|^2 < \infty.\end{equation}
These are Hilbert spaces with inner product
\begin{equation} \llangle \eta , u \rrangle_m = |Q| \sum_{\bm{k}\in\mathbb{Z}^2} \left(1 + |\bm{\tilde{k}}| \right)^{2m} \eta_{\bm{k}} u_{-\bm{k}}.\end{equation}
We have the following properties for the non-local operator $\mathcal{R}$ defined by (\ref{operatorRdefinition}) (which are all trivial to prove from the definition and symbol):
\begin{enumerate}[(i)]
\item {\textrm{ }} $\mathcal{R}$ commutes with derivatives
\item {\textrm{ }} $\mathcal{R}$ is self adjoint on $H^m$
\item {\textrm{ }} We have that 
\begin{equation} \| \mathcal{R}(\eta) \|_{m} \leq \| \eta \|_{{m+1}} \end{equation}
and also
\begin{equation} | \mathcal{R}(\eta) |_2^2 = | \eta_x |_2^2 + | \eta_y |_2^2 \end{equation}
\end{enumerate}
In fact, $\mathcal{R}$ is an isometry of homogeneous Sobolev spaces from $\tilde{H}_{\textrm{per}}^m(Q)$ to $\tilde{H}_{\textrm{per}}^{m+1}(Q)$, where $\tilde{H}_{\textrm{per}}^m(Q)$ has norm
\begin{equation}|Q|^{1/2}\left(\sum_{\bm{k}\in\mathbb{Z}^2}|\bm{\tilde{k}}|^{2m} |\eta_{\bm{k}}|^2\right)^{1/2}.\end{equation}

\section{Estimates for numerics \label{appendixnumericsposdef}} 

We now derive a condition on $c$ to ensure that the operator $\mathcal{A}$ defined in (\ref{mathcalABdefn}) is positive definite. Firstly by Cauchy-Schwarz and integration by parts,
\begin{align}\langle \mathcal{A} \eta, \eta  \rangle_2  = & - (\beta - 1) | \eta_{x} |_2^2 + | \eta_{y} |_2^2 - \gamma \langle  \mathcal{R}(\eta) , \Delta \eta \rangle_2 + |\Delta \eta|_2^2  + c |\eta|_2^2  \label{Ainnerproduct1}\\
\geq & - |\beta - 1| | \eta_{x} |_2^2 - | \eta_{y} |_2^2 - \gamma | \Delta \eta |_2 | \mathcal{R}(\eta) |_2 + \frac{1}{2} |\Delta \eta|_2^2 + \frac{1}{2} |\eta_{xx}|_2^2 + \frac{1}{2} |\eta_{yy}|_2^2 + c |\eta|_2^2.\nonumber 
\end{align}
From the properties of $\mathcal{R}$ in Appendix \ref{PropertiesofRappendix} we have
\begin{equation}| \mathcal{R}(\eta) |_2 = \sqrt{ | \eta_x |_2^2 + | \eta_y |_2^2} \leq | \eta_x |_2 + | \eta_y |_2,\end{equation}
and Young's inequality gives
\begin{subeqnarray}
\gdef\thesubequation{\theequation \textit{a,b}} |\eta_x|_2^2 \leq \frac{1}{2\epsilon_1} |\eta|_2^2 + \frac{\epsilon_1}{2} | \eta_{xx} |_2^2,\;\quad\quad &\displaystyle{ |\eta_y|_2^2 \leq \frac{1}{2\epsilon_2} |\eta|_2^2 + \frac{\epsilon_2}{2} | \eta_{yy} |_2^2,} \\
\gdef\thesubequation{\theequation \textit{c,d}} \gamma | \Delta \eta |_2 | \eta_x |_2 \leq \frac{\epsilon_3}{2} | \Delta \eta |_2^2 + \frac{\gamma^2}{2\epsilon_3} | \eta_x |_2^2,\quad & \displaystyle{\gamma | \Delta \eta |_2 | \eta_y |_2 \leq \frac{\epsilon_4}{2} | \Delta \eta |_2^2 + \frac{\gamma^2}{2\epsilon_4} | \eta_y |_2^2,}
\end{subeqnarray}
for any $\epsilon_1,\epsilon_2, \epsilon_3, \epsilon_4 > 0$. Then using these with (\ref{Ainnerproduct1}) yields
\begin{align}\nonumber \langle \mathcal{A} \eta, \eta  \rangle_2  \geq & \left( c - \frac{|\beta - 1|}{2\epsilon_1}  - \frac{1}{2\epsilon_2} - \frac{\gamma^2}{4\epsilon_1 \epsilon_3} - \frac{\gamma^2}{4\epsilon_2 \epsilon_4} \right)|\eta|_2^2 + \left(\frac{1}{2} -\frac{|\beta - 1|\epsilon_1}{2} - \frac{\gamma^2 \epsilon_1}{4 \epsilon_3} \right)|\eta_{xx}|_2^2\\
& + \left( \frac{1}{2} - \frac{\epsilon_2}{2} - \frac{\gamma^2 \epsilon_2}{4 \epsilon_4}\right)|\eta_{yy}|_2^2 + \left(\frac{1}{2} - \frac{\epsilon_3}{2} - \frac{\epsilon_4}{2} \right)|\Delta \eta |_2^2.
\end{align}
Taking
\begin{equation}\epsilon_1 = \frac{1}{\left(|\beta - 1| + \gamma^2\right)}, \quad \epsilon_2 = \frac{1}{\left(1 + \gamma^2\right)} , \quad \epsilon_3 = \epsilon_4 = \frac{1}{2},\end{equation}
ensures that all the brackets preceding norms of derivative terms are zero. So to ensure that $\mathcal{A}$ is positive definite it is sufficient to take
\begin{equation}c > \frac{|\beta - 1|}{2\epsilon_1}  + \frac{1}{2\epsilon_2} + \frac{\gamma^2}{4\epsilon_1 \epsilon_3} + \frac{\gamma^2}{4\epsilon_2 \epsilon_4} = \frac{1}{2} \left[ \left(|\beta - 1|+ \gamma^2\right)^2 + \left(1 + \gamma^2\right)^2 \right]. \end{equation}

RJT acknowledges the support of a PhD scholarship by EPSRC. The work of DTP was supported by EPSRC grants EP/K041134 and EP/L020564, and the work of GAP was supported by EPSRC grants EP/L020564, EP/L025159 and EP/L024926.

\bibliographystyle{unsrt}

\bibliography{ElecThinFilmBib}

\end{document}